\documentclass[manuscript, sigconf]{acmart}

\AtBeginDocument{%
  \providecommand\BibTeX{{%
    \normalfont B\kern-0.5em{\scshape i\kern-0.25em b}\kern-0.8em\TeX}}}

\copyrightyear{2025}
\acmYear{2025}
\setcctype{by}
\acmConference[CHI '25]{CHI Conference on Human Factors in Computing Systems}{April 26-May 1, 2025}{Yokohama, Japan}
\acmBooktitle{CHI Conference on Human Factors in Computing Systems (CHI '25), April 26-May 1, 2025, Yokohama, Japan}\acmDOI{10.1145/3706598.3713139}
\acmISBN{979-8-4007-1394-1/25/04}

\usepackage{graphicx}  
\usepackage{url} 
\usepackage{subcaption}
\usepackage{multirow}
\usepackage{listings}

\usepackage[commandnameprefix=ifneeded, final]{changes}

\newcommand{\canvil}{\textsc{Canvil}}
\newcommand{\textcanvil}[1]{\textit{\color{violet}#1}}
\graphicspath{{figures/}{pictures/}{images/}{./}}

\begin{document}

\title[\canvil{} for Designerly Adaptation]{\canvil{}: Designerly Adaptation for LLM-Powered \\User Experiences}

\author{K. J. Kevin Feng}
\authornote{Work done while at Microsoft Research.}
\affiliation{%
  \institution{University of Washington}
  \city{Seattle, WA}
  \country{USA}}
\email{kjfeng@uw.edu}

\author{Q. Vera Liao}
\affiliation{%
  \institution{Microsoft Research}
  \city{Montréal, QC}
  \country{Canada}}
\email{veraliao@microsoft.com}

\author{Ziang Xiao}
\authornotemark[1]
\affiliation{%
  \institution{Johns Hopkins University}
  \city{Baltimore, MD}
  \country{USA}}
\email{ziang.xiao@jhu.edu}

\author{Jennifer Wortman Vaughan}
\affiliation{%
  \institution{Microsoft Research}
  \city{New York City, NY}
  \country{USA}}
\email{jenn@microsoft.com}

\author{Amy X. Zhang}
\affiliation{%
  \institution{University of Washington}
  \city{Seattle, WA}
  \country{USA}}
\email{axz@cs.uw.edu}

\author{David W. McDonald}
\affiliation{%
  \institution{University of Washington}
  \city{Seattle, WA}
  \country{USA}}
\email{dwmc@uw.edu}

\renewcommand{\shortauthors}{Feng, et al.}

\begin{teaserfigure}
  \centering
  \includegraphics[width=\textwidth]{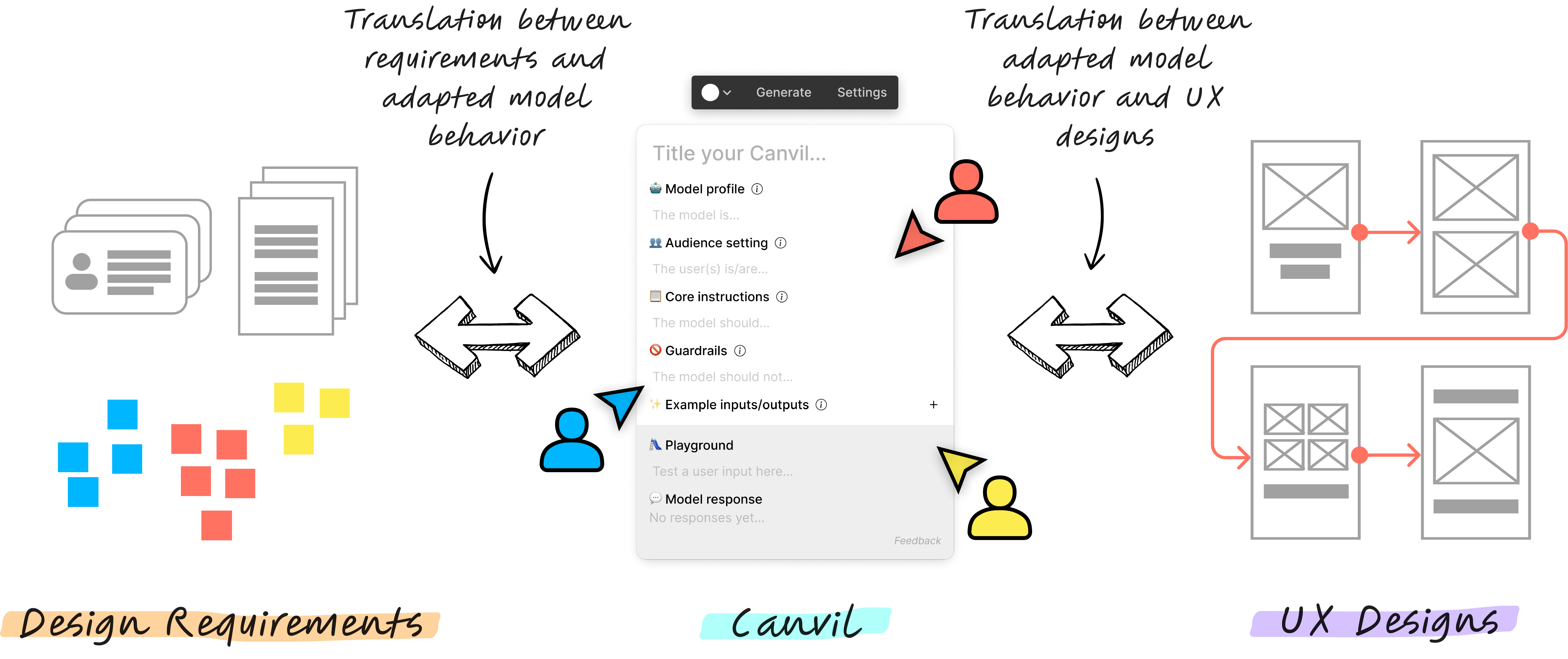}
  \caption{We propose \textit{designerly adaptation} as a means of exploring LLMs as a design material, enabling a two-way translation between design requirements and UX designs for LLM-powered applications. We then present \canvil{}, a Figma widget that operationalizes designerly adaptation in design canvases. By using \canvil{} as a probe, we found that designerly adaptation surfaced LLM behavior that allowed designers to co-evolve design requirements and UX designs, and thoughtfully craft end-user interaction with LLMs.}
  \label{fig:teaser}
  \Description{Diagram showing the integration of design requirements and UX designs through a tool called Canvil. On the left, 'Design Requirements' are represented by icons of documents and sticky notes. In the center, 'Canvil' is depicted as a user interface with settings for model profiles, audience settings, guardrails, and a playground. Three users are shown interacting with Canvil, indicating real-time collaboration. On the right, 'UX Designs' are represented by wireframe sketches of a digital interface, connected by red lines to indicate a flow or iteration process. Canvil facilitates a two-way translation between design requirements and UX designs.}
\end{teaserfigure}

\begin{abstract}
  Advancements in large language models (LLMs) are sparking a proliferation of LLM-powered user experiences (UX). In product teams, designers often craft UX to meet user needs, but it is unclear how they engage with LLMs as a novel design material. Through a formative study with 12 designers, we find that designers seek a translational process that enables design requirements to shape and be shaped by LLM behavior, \replaced{motivating a need for}{and propose} \textit{designerly adaptation} to facilitate this translation. We then built \canvil{}, a Figma widget that operationalizes designerly adaptation. We used \canvil{} as a \deleted{technology }probe \added{to study designerly adaptation} in a group-based design study (6 groups, $N=17$), finding that designers constructively iterated on both adaptation approaches and interface designs to enhance end-user interaction with LLMs. Furthermore, designers identified promising collaborative workflows for designerly adaptation. Our work \replaced{opens new avenues for processes and tools that}{demonstrates the promises of designerly adaptation to} foreground designers' human-centered expertise when developing LLM-powered applications.
\end{abstract}



\maketitle

\section{Introduction}
A paradigm shift is underway for integrating artificial intelligence (AI) capabilities into everyday user-facing technologies. Large pre-trained AI models, most notably large language models (LLMs), have versatile natural language capabilities that unlock novel interactive techniques and interfaces for more intuitive and customizable user experiences across a wide spectrum of applications \cite{suh2023sensecape, angert2023spellburst, wu2022ai, parnin2023building}. However, these promises also come with numerous concerns. Integrating LLMs into a domain without careful consideration of the user contexts surrounding model use and implementation of appropriate guardrails may result in user experiences that perpetuate societal biases \cite{santurkar2023whose}, threaten users' sense of well-being \cite{snapchat-my-ai, bingchat-nyt}, or otherwise do harm \cite{selbst2019abstraction, pang2023anticipate, weidinger2022taxonomy, chan2023harms, shelby2023sociotechnical}. 

As technology development practitioners, designers\footnote{We consider designers to be anyone who has an active, hands-on role in designing the user experience of a product and/or feature. This includes job titles such as UX designer/researcher, content designer, and product designer.} are uniquely positioned to mitigate these concerns \cite{feng2023addressing, subramonyam2021protoai, liao2023designerly, yang2018material, yildirim2022experienced} \added{and operationalize growing calls for human-centered AI} \cite{shneiderman2022human, landay-hai}. Designers' work often involves aligning technological capabilities with user needs, such that the technology addresses (or makes progress towards addressing) pain points identified in user research \cite{feng2023addressing, subramonyam2022leaky}. Designers are trained in human-centered design methods that allow them to understand users and usage contexts, prototype potential solutions with relevant technology, and iterate on those solutions based on their understanding of users or user feedback \cite{double-diamond}. Yet, prior work has shown that designers face diverse procedural challenges (e.g., difficulties collaborating with the engineering teams training the models \cite{subramonyam2022leaky, subramonyam2021towards}) and instrumental challenges (e.g, lacking means to effectively work with the models \cite{feng2023addressing, yang2018material, subramonyam2021protoai}) when working with AI \cite{yang2020difficult, subramonyam2021towards}. In efforts to address these challenges, researchers have situated AI as a design material to highlight its material properties---including the technology's capabilities, limitations, and adaptability \cite{leonardi2012materiality}---for designers to better understand and apply AI in the context of their design problems \cite{yang2018material, dove2017material, feng2023ux}. A good \textit{designerly understanding} of AI can help designers ideate on new AI-powered design ideas, mitigate AI’s varying impact for different user scenarios, collaborate with design and non-design practitioners, and reinforce user-centered perspectives amongst the team and throughout the product development cycle \cite{liao2023designerly,wang2023rai}. 

The advent of powerful pre-trained LLMs introduces new opportunities for engaging with AI as a design material. First, adaptability emerges as a key materialistic property of LLMs---these models are responsive to adaptation via fine-tuning \cite{dodge2020fine, hu2021lora, ouyang2022rlhf, bai2022constitutional} and prompt-based methods \cite{openai-system-prompt, wu2022promptchainer, OpenAI2023GPT4TR}. In fact, due to resource-intensive training of LLMs, it has become a standard practice for individual developers and development teams to adapt ``base'' LLMs from a small handful of providers (e.g., Anthropic, Google, OpenAI) instead of training custom in-house models \cite{parnin2023building}. Moreover, natural language interaction and adaptation democratizes AI experimentation for practitioners---including designers---traditionally excluded from AI conversations due to limitations in technical expertise \cite{liao2023designerly, petridis2023promptinfuser}. Despite this, there has been limited exploration of designers' interaction with---let alone adaptation of---LLMs in practice as they contribute to LLM-powered features and products. \looseness=-1

In this paper, we first conduct a formative interview study with 12 designers experienced in designing features and products that use LLMs to better understand their workflows and desiderata when crafting LLM-powered UX. From our interviews, we identify a need for a translational process that allow design requirements (e.g., findings from user research, product goals) to directly shape model behavior, and vice versa. Drawing from Cross' notion of \textit{design codes}---systematic ways of representing, transmitting, and translating design knowledge \cite{cross1982designerly}---we propose \textit{designerly adaptation for LLMs} (henceforth ``designerly adaptation'') as a new design code to facilitate this two-way translation. We operationalize designerly adaptation by building \canvil{}, a Figma\footnote{Figma is a popular collaborative design tool: https://www.figma.com/.} widget that supports structured authoring of model behavioral specifications based on design requirements to adapt LLMs, testing of adapted models on diverse user inputs, and integration of model outputs into interface designs. \replaced{We then use \canvil{} as a technology probe to study designerly adaptation with 6 groups with a total of 17 designers. We demonstrate the effectiveness of our probe by collecting concrete, empirical insights into how designers engage in designerly adaptation through \canvil{}. Specifically, we}{To investigate whether and how designerly adaptation can assist in crafting LLM-powered UX in practice, we used \canvil{} as a technology probe in a task-based design study with 6 groups of 17 designers. We}
find that \deleted{through using \canvil{}, }designers iteratively translated diverse design requirements into LLM behavioral specifications. This allowed them to simultaneously adapt their designs to enhance user interaction with LLMs, while setting or refining behavioral constraints for the model using their designs. Furthermore, designers recognize promises in collaboratively engaging with designerly adaptation to share resources and knowledge with design and non-design stakeholders. They are optimistic about integrating designerly adaptation into their own workflows and note additional procedural and instrumental questions to address in practice. Concretely, our work makes the following contributions:
\begin{itemize}
    \item \replaced{Designerly adaptation---a process bridging material exploration of LLMs with design processes---motivated by insights from a formative study with designers who craft LLM-powered UX}{Insights from a formative study with designers experienced in crafting LLM-powered user experiences, from which we motivate designerly adaptation---a practice bridging material exploration of LLMs with design processes.}.
    \item \canvil{}, a technology probe \added{to operationalize designerly adaptation} in the form of a Figma widget \added{and allow us to study it in practice}.
    \item \replaced{Empirical learnings uncovered by \canvil{}}{Insights} about how designers engage in designerly adaptation in a task-based design study.
    \item A discussion of our learnings' implications on design practice and beyond, including a proposed workflow for designerly adaptation to orient future work and a reflection of LLMs' sociomaterial implications on collaborative practices within product teams.
\end{itemize}

\section{Related Work}

\subsection{AI as a Design Material}
Robles and Wiberg argue that recent advances in computational technologies bring about a ``material turn''---a transformation within interaction design that allows for the shared use of material metaphors (e.g., flexibility) across physical and digital worlds \cite{robles2010material}. 
Indeed, prior works have discussed AI as a \textit{design material} \cite{dove2017material, yang2018material, yang2020difficult, benjamin2021material, luciani2018material, liao2023designerly, feng2023addressing, feng2023understanding, yildirim2022experienced}, and highlighted why AI's materiality can make it uniquely difficult to design with \cite{yang2020difficult, subramonyam2021towards, liao2023designerly, dove2017material, benjamin2021material}.  AI is often treated as a black box to non-technical stakeholders such as some designers, making it challenging to tune user interactions to often unpredictable and complex model behavior \cite{benjamin2021material, yang2018material}. In addition, AI's technical abstractions are often divorced from concepts designers are familiar with \cite{subramonyam2021towards}, and designers consequently struggle with creatively using or manipulating the material to generate design solutions \cite{liao2023designerly, feng2023addressing}. AI models are also non-deterministic and fluid in nature---they may evolve with new data or user input, and can be intentionally steered towards desirable behaviors with choices of data, algorithms, parameters, and so on \cite{yang2020difficult}. Without a concrete material understanding to begin with, designers face challenges grasping the nature of these uncertainties \cite{yang2018material, subramonyam2021protoai} or the opportunities to shape the design materials for desirable UX \cite{liao2023designerly}.
Yang et al. \cite{yang2020difficult} showed that these challenges of working with AI as a design material persist through the entire double-diamond design process---from identifying the right user problem to be solved by AI to designing the right UX to solve the user problem. \added{These challenges pose barriers to developing human-centered AI---expanding the algorithm-centered scope of AI and applying methods from human-centered design to ensure that the development of the technology can better serve human needs} \cite{shneiderman2022human, shneiderman2020human, capel2023human, riedl2019human, landay-hai}.

Researchers and practitioners have developed processes and tools to alleviate some of these challenges. ProtoAI \cite{subramonyam2021protoai} combines exploration of models with UI prototyping, while advocating for designers' active shaping of the AI design material (e.g., choosing models and setting parameters) by user needs. Feng et al. \cite{feng2023addressing} found that hands-on ``fabrication'' of the design material through a UI-based model training tool bolstered understanding and connection between AI properties and UX goals. fAIlurenotes \cite{moore2023failurenotes} is a failure analysis tool for computer vision models to support designers in understanding AI limitations across user groups and scenarios. Other efforts include process models \cite{subramonyam2021towards} and ``leaky abstractions'' \cite{subramonyam2022leaky} that facilitate collaboration between designers and model developers, and human-AI design guidelines \cite{apple-guidelines, pair-guidebook, hax-guidelines}.

Recent advances in LLMs can simultaneously address and exacerbate some of the aforementioned challenges. The barrier to tinkering with AI has significantly lowered thanks to the use of natural language as a primary mode of interaction and easily accessible tools such as ChatGPT. Concerns, however, have also arisen over the lack of transparency and controllability in LLMs due to their complex technical architectures \cite{liao2023ai}. Yet, because of LLMs' powerful capabilities, there is significant interest in exploring their integration into user-facing applications  \cite{landay-hai, parnin2023building}; as such, designers should be prepared to work with them as a design material \cite{kulkarni2023word}. \replaced{While past work advocated for a better ``designerly understanding'' upon studying how to support designers in working with models \protect{\cite{liao2023designerly}}, our work specifically explores involving designers in the practice of adapting LLMs. Adaptation and understanding can be related and synergistic, but our focus is on adaptation---a common practice in product teams \protect{\cite{parnin2023building}}---and how it can support designers to craft LLM-powered UX.}{Despite this, we have yet to understand designers' current approaches and desiderata when crafting LLM-powered user experiences. Our work contributes both empirical knowledge and tooling via a technology probe to this space.}

\subsection{Adaptation of Large Language Models}
\label{s:rw-llm-adaptation}
A fundamental property of LLMs not present in their smaller predecessors is the ease with which their behavior can be adapted \cite{brown2020language, OpenAI2023GPT4TR}. While the breadth of LLMs' out-of-the-box capabilities may seem impressive, researchers have recognized the importance of adapting LLMs for enhanced performance under specific domains and tasks \cite{dodge2020fine, hu2021lora, wang2023rai, wu2022ai, wu2022promptchainer}, and aligning model behavior with human preferences and values \cite{ouyang2022rlhf, bai2022constitutional, feng2023case}. Adaptation has thus been a topic of interest to both the AI and HCI communities \cite{lee2022coauthor}.

Adaptation may take on many forms \added{and can be performed by different stakeholders}. \replaced{AI engineers and data scientists may fine-tune a pre-trained model by using a task-specific dataset to train additional layer(s) of the neural network}{A pre-trained model may undergo \textit{fine-tuning}, a process by which additional layer(s) of the neural network are trained on a task-specific dataset} \cite{dodge2020fine}. Popular \added{technical} approaches to fine-tuning include instruction tuning \cite{wei2021finetuned, ouyang2022rlhf}, reinforcement learning with human feedback (RLHF) \cite{ouyang2022rlhf, christiano2017deep}, and direct preference optimization (DPO) \cite{rafailov2023direct}. More computationally efficient variants, such as low-rank adaptation (LoRA) \cite{hu2021lora}, have also garnered attention.

\added{Besides technical stakeholders, HCI has long observed that end users often adapt their technology---and also learn to perform this adaptation over time---in a process Mackay calls \textit{co-adaptation}} \cite{mackay1990users, mackay2024parasitic}. \replaced{End-user adaptation of LLMs is made accessible by methods that modify model behavior without modifying the model itself, such as system prompting}{Adaptation can also occur without modifying the model itself via system prompting} \cite{msft-system, wei2023jailbroken, OpenAI2023GPT4TR, salewski2023context, deshpande2023toxicity}. 
Different from one-off user prompting, system prompting applies to all individual user inputs for how the model should behave (e.g., ``\textit{always respond in a concise manner}''), often targeting an application domain. 
For instance, a line of work explored instructing the model to behave with a certain ``persona'' to elicit desirable or adversarial model behaviors \cite{wei2023jailbroken, cheng-etal-2023-marked, salewski2023context, wolf2023fundamental, deshpande2023toxicity}. However, system prompts can be challenging to author \cite{zamfirescu2023johnny}, and there is not yet an established ``gold standard'' for prompt authoring, system or otherwise. \added{Co-adaptation is a two-way street \protect{\cite{mackay2024parasitic}}, and we do not yet have a systematic understanding of how to adapt models even though models can quickly adapt to our prompts.} Researchers and practitioners have thus attempted to derive useful prompting formats based on empirical exploration; these include in-context learning (i.e., by providing desired input-output examples) \cite{brown2020language, wu2023scattershot, zhao2021calibrate}, chain-of-thought reasoning \cite{wei2022chain}, and instruction-following \cite{kaddour2023challenges, ouyang2022training}. Industry recommendations have also emerged for system prompts, encouraging the specification of elements such as context specification (e.g., ``\textit{You are Yoda from Star Wars}''), task definition (e.g., ``\textit{You respond to every user input as Yoda and assume the user is a Padawan}''), and safety guardrails (e.g., ``\textit{If the user requests inappropriate or offensive responses, you must respectfully decline with a wise Yoda saying}'') \cite{msft-system}.  

In our work, we examine and support \textit{designerly adaptation} of LLMs---a means of empowering designers to explore LLMs as a design material through model adaptation. \added{Unlike many technical adaptation methods, designerly adaptation contributes to human-centered AI by drawing from designers' workflows and expertise. This focuses efforts to ensure that AI-powered applications satisfy user needs and effectively co-adapt with us as they become more pervasive.} While we draw from recent developments in and recommended best practices for accessible model adaptation via natural language, we acknowledge that these practices may shift over time \replaced{as we co-adapt to this new technology. Using}{and using} currently available techniques is just one possible instantiation of designerly adaptation.

\subsection{Interactive Tools for Steering AI Behavior}

Literature at the intersection of HCI and AI has introduced a wide range of interactive techniques and tools to aid humans in training and adapting AI models, ranging from ones supporting data scientists to perform data wrangling \cite{gortler2022neo, wang2022timbertrek, kandel2011wrangler, liu2020boba, wongsuphasawat2015voyager}, model training and evaluation \cite{bauerle2022symphony, cabrera2023zeno, amershi2015modeltracker, ren2016squares, ms-aml, google-aml, ibm-aml}, managing model iterations \cite{hohman2020iteration, robertson2023angler}, and so on, to those allowing for ``human-in-the-loop'' paradigms at various stages of the model development pipeline, including data annotation \cite{smith2018closing, settles2012active, chao2010transparent, monarch2021human, raghavan2006active}, output correction \cite{branson2010visual, boden2021human}, and integration testing \cite{chen2022hint}, and explainability.

There has also been interest in ``democratizing AI'' for domain experts or practitioners without formal technical training to steer model behaviors. Interactive machine learning (iML) \cite{fails2003iml} is a field responding to this interest by advocating for interactive and incremental model steering through intuitive interfaces and tightly coupled input-evaluation feedback loops.  
For instance, interfaces for transfer learning \cite{mishra2021transfer} have enabled non-technical audiences to transfer learned representations in a model to a domain-specific task. Tools have also encouraged exploratory tinkering of AI models through visual drag-and-drop UIs \cite{carney2020teachable, liner, lobe, du2023rapsai}. Teachable Machine \cite{carney2020teachable} is one such tool that allows users to train models for image, video, and audio classification. Rapsai \cite{du2023rapsai} is a visual programming pipeline for rapidly prototyping AI-powered multimedia experiences such as video editors. In an era where formal knowledge about AI is limited to technical experts, these tools also serve to demystify AI for everyday users. 

With the onset of LLMs, barriers to experimenting with AI have lowered even further, paving the path for a new generation of interactive AI tools. Sandbox environments such as OpenAI's playground \cite{openai-playground} require no prerequisites besides a grasp of natural language to prompt the model and steer model behavior. However, relying on unstructured natural language alone can be daunting and ineffective. Many tools have risen to the challenge to support prompt engineering with more structured tinkering. Prompt chaining is one such approach \cite{wu2022ai, wu2022promptchainer, Arawjo_2023, suh2023sensecape}, by which users can use a node-based visual editor to write prompts for simple subtasks and assemble them to solve a larger, more complex task. 
PromptMaker \cite{jiang2022promptmaker} and MakerSuite \cite{makersuite} allow the user to rapidly explore variable-infused prompts and few-shot prompting, while ScatterShot \cite{wu2023scattershot} helps specifically with curating few-shot prompting examples. 

Despite advancements in tooling, support for designers to work with LLMs as an adaptable design material remains limited. Domain-agnostic tools for tinkering with models may not be well-integrated into design workflows, a primary consideration for designers when deciding whether to adopt those tools \cite{feng2023understanding}. Tools that offer integration with design environments (e.g., PromptInfuser \cite{petridis2023promptinfuser, petridis2023promptinfuser-lbw}) support prototyping with LLMs, but not necessarily deeper adaptation of model behavior. Our work situates interactive adaptation within familiar design environments and processes.

\section{Formative Study}
Working with AI, especially LLMs, is an emerging practice in the field of UX \cite{yang2020difficult, yildirim2023guidebook, feng2023ux, subramonyam2021towards}. We conducted a formative interview study to understand designers' experiences working on LLM-powered products and features amidst industry-wide shifts towards LLMs in 2023 \cite{parnin2023building} as an initial step in our research. 

\subsection{Method and Participants}

We conducted 30-minute 1:1 virtual interviews with 12 designers at a large international technology company where LLM-powered features and products are actively explored. \added{Our interviews were not initially focused on model adaptation, let alone designing a probe for adaptation. Instead, our goal was to more deeply understand designers' workflows and pain points they face when crafting LLM-powered UX. Thus, we opted for interviews as our method of choice.} All interviews were conducted in June and July 2023. At the time of the study, all participants were working on products or services that leveraged LLMs in some capacity; LLM application areas spanned conversational search, question-answering (QA) on domain-specific data, recommendation, text editing and generation, and creativity support tools. Participant details can be found in Table \ref{t:participants-1} of Appendix \ref{a:formative}.

Our interviews were semi-structured and revolved around the following topics:
\begin{itemize}
    \item \textbf{Awareness:} to what extent are designers aware of LLMs' capabilities, limitations, and specifications in the context of their product(s)?
    \item \textbf{Involvement:} to what extent are designers involved in discussions or activities that shape where and how an LLM is used in their product(s)?
    \item \textbf{Desiderata:} what do designers desire when crafting LLM-powered UX, with regards to both processes and tools?
\end{itemize}

Each participant received a \$25 USD gift card for their participation. All interviews were recorded and transcribed. Our study was reviewed and approved by the company's internal IRB. 

The first author performed an inductive qualitative analysis of the interview data \added{using the qualitative coding platform Marvin}\footnote{https://heymarvin.com/}. This process started with an open coding round in which initial codes were generated, followed by two subsequent rounds of axial coding in which codes were synthesized and merged into \replaced{(yet-to-be-named) groups, in line with Braun and Clarke's reflexive thematic analysis \cite{braun2019reflecting}}{higher-level themes}. \added{These groups were reviewed and named through an iterative process that involved the disassembling and merging of groups; the named groups became initial themes.} The codes and themes were discussed with research team members at weekly meetings. Other research team members also offered supporting and contrasting perspectives \added{(e.g., (dis)agreement on main themes, interpretation of new themes)} by writing their own analytical memos, which were also discussed as a team. \added{Any disagreements were resolved through team meetings and asynchronous discussion via comments on the analytical memos.}

\subsection{Findings}

\subsubsection{Adaptation was seen as a central materialistic property of LLMs}
\label{s:formative-adaptation-materialistic}

Participants recognized that products delivering compelling, robust UX were not powered by out-of-the-box ``base'' LLMs, but required adaptation based on user needs and contexts. P5 gave an example where an out-of-the-box model inappropriately prompted the user for a riddle in a workplace setting: \textit{``[The experience] is not quite right, cause it's in [workplace software] and it's like, tell me a riddle. I'm at work!''} P4 and P7 both stated a design goal for their products (in conversational search and domain-specific question-answering, respectively) is to accommodate non-technical users, which can be achieved if the model can \textit{``tailor the language depending on technical ability, and gradually introduce [users] to more complicated concepts and terminology''} [P7]. P2 shared that their team settled on using three adapted versions of the same model to customize UX within their product: \textit{``There are actually three incarnations of the same model, each with slightly different parameters [...] based on the prompt that users give, we select which of those 3 incarnations we wanna use.''}

In their design work, participants appreciated the ability to steer an LLM with natural language to envision more flexible and adaptable AI-powered UX. Some were familiar with how lightweight adaptation techniques, such as writing system prompts, can be used to specify model behaviors based on design specifications. P12 discussed an example of how this may work in an entertainment system setting: 
\begin{quote}
    \textit{``You might want to generate enthusiasm more, right? So the LLM that goes over there might have a UX layer that feels different. The system prompt can say, remember, you're a machine that wants to get users enthused.''} [P12]
\end{quote}

P12's approach points to an \textbf{important distinction in outcome evaluation for when a model is adapted for design purposes versus conventional prompt engineering:} the end solution to a design problem is not a prompt, but a user experience. In P12's case, they opted for the model to act with enthusiasm in their product. Whether the model can generate enthusiastic responses and whether the model delivers a desirable user experience within the product are two different (albeit potentially related) questions. The latter likely also depends on design choices made elsewhere in the product and should be evaluated through user studies rather than model performance metrics.

\subsubsection{Designers envisioned constraints and desiderata for model behavior, but were unable to directly apply them}
\label{s:formative-involvement-normative}
Participants often proposed constraints and desiderata for model behavior in their design work. For example, P11 explained that in their product, designers set character limits for LLM-generated summaries in the UI, while P5 and P12 both created ``personalities'' for the LLM in their products. However, participants rarely had the opportunity to tinker with and apply their desiderata and constrints themselves, resulting in overly cumbersome workflows. P11 shared that they relied on engineers to adapt on their behalf: \textit{``I was asking what the difference in summary looks like with 50, 100, and 150 characters. Then [the engineers] would go back and test it and then they would just share the results, like here's what 50 characters looks like. Here's what 100 looks like.''} P12 experienced a similar procedure and stated that the feedback loop can take as long as a couple weeks. Ideally, they wanted the experimentation to happen \textit{``in real time.''} In general, designers were typically not involved in model adaptation decisions. Some were unsure of how or by whom adaptation is performed and suspected that it was engineers: \textit{``I would say [adaptation]'s probably something that was decided by the engineering team [...] prompt engineering is more so on the [technical] side of things.''} [P2]. 

Because model behavior and UX are inextricably linked, many participants desired more involvement in steering models directly, particularly towards UX goals. P4 shares: \textit{``Who better to involve in this process than people whose job it is to think deeply about [the UX]? At the very least, system prompts shouldn't be written without an understanding of what the end UX goals are.''} P9 agreed, saying that designers can offer strategic contributions with user-centered thinking: \textit{``It's super important for designers to think through ideas and make it intuitive for users and help come up with compelling [user] scenarios [...] I think design has a bigger opportunity to have a seat at the table strategy-wise.''} On a higher level, P12 pointed out that model adaptation can be a contemporary extension of efforts around crafting product voice and tone, which designers are already familiar with: \textit{``UX designers and content designers are very attuned to and have pretty much owned the story around voice and tone of products, and have for years and years.''} 

We see strong evidence showing that designers are \textit{uniquely positioned} to contribute to adaptation through their user-centered lens. Although they may primarily adapt models via prompt-based methods, \textbf{their contributions extend far beyond prompt engineering.} First, they define a key prerequisite for prompt engineering: the UX goals which the model should be steered towards and evaluated against. This is not a one-way process---these goals may be modified based on hands-on experimentation with the model. However, we did not observe a two-way process in practice as designers were often not involved in adaptation themselves. Second, they can surface additional constraints and desiderata for prompt engineered models through UX evaluation and testing. Indeed, P10 caught their model's inability to properly recognize some acronyms in their product during testing and devised a solution with their team: \textit{``I found that [the model] doesn't know what to do with the acronyms so we floated the idea of having glossary of industry jargon and acronyms [in the system prompt].''} In sum, designers' user-centered insights help improve LLM-powered UX, addressing crucial decisions in both goal-setting \textit{before} prompt engineering and UX evaluation \textit{after} the fact.

\subsubsection{Designers did not find sufficient support in current LLM tools and resources} 
\label{s:formative-lacking-support}
Overall, participants felt limited by current tools and resources for tinkering with LLMs within the design process. Prior work found that the inability to directly access the models and experiment with their capabilities and limitations is a primary challenge in the design process for AI-powered UX \cite{yang2020difficult, subramonyam2022leaky}. This was reflected in our findings as well---in the few cases where designers had hands-on access to model tinkering, designers recalled the experience being highly valuable. P2 said they could much more clearly \textit{``see or gauge the power of the language model,''} while P9 tinkered with a model they had early access to and recalled that \textit{``you can identify some gaps [in capability] right off the bat.''} However, most did not have access to tinkering and found this frustrating. Designers either had to keep burdening the engineers or wait until a test version of the product is launched. P3 shared that \textit{``The only way we would have [to tinker] is using the [product] online and play around with what [the team] did.''} Similarly, P10 felt limited by this workflow: \textit{``You want to be able to play with [the model] yourself and understand what the UX is like. And I can't play with it.}'' 

Some designers took the initiative to seek out new tools to tinker with LLMs. Designers welcomed the easy access to free LLM chatbots such as ChatGPT, but did not feel like they were well-integrated into their design workflows. Many tried a variety of tools---mostly playground interfaces for prompt engineering---without much success. Among others, P1 did not find these interfaces to be designer-friendly: \textit{``[The tools are] still kind of technical, a lot of [designers] don't realize how many parameters work. Like how does temperature work?''} Others wanted these tools to leverage the same metaphors as already-familiar design tools, such as \textit{``a UX library with some specific patterns you can use''} [P4] or \textit{``accessing content right away in a [design system]''} [P6].

Our findings \textbf{reveal fundamental differences in tooling needs} between designers who adapt LLMs for UX design purposes and prompt engineers. For designers, the model is a design material through which they explore design solutions. That is, unlike prompt engineering, the material itself is not the final product---the UX is. The UX does not stop nor end with model behavior. For example, designers will also need to consider user journeys leading up to the LLM interaction, and subsequent user interactions with the LLM output as well as other UI components  Designers work with many different types of materials within the broader design process, and typically have adequate tools to understand and manipulate these materials. For example, designers can easily draft wireframes and create animations in modern design tools such as Figma. When it comes to understanding and manipulating LLMs, designers are much less well-supported. Prompt engineering tools do not satisfy designers' needs because \textit{they help craft prompts rather than UX and they do not fit into existing design processes}.

\subsection{Summary: The Case For Designerly Adaptation}
\label{s:designerly-adaptation-def}

In his 1982 essay, Nigel Cross characterized the uniqueness of epistemological inquiry in the field of design by outlining several aspects of \textit{designerly way of knowing} \cite{cross1982designerly}. One key aspect was the use of \textit{design codes}. To Cross, design codes offer a two-way translation between abstract design requirements and concrete designed objects---that is, codes represent the \textit{entanglement of a design process with its design materials}. The process generates requirements that, when applied to the materials, results in a designed object, while the properties of the material may simultaneously shape the process used to work with it. In software development, coding is a process that simultaneously implements system specifications and allows for refinement of the specifications based on implemented behavior. Design codes serve a similar purpose for design requirements and designed objects.

For some designers in our formative study, LLM prompts provided an accessible means of expressing design requirements in natural language to adapt model behavior. However, prompting alone did not offer the entanglement needed to shape a designed object (Section \ref{s:formative-involvement-normative}), nor did it allow LLMs' material properties to influence designers' practices due to prompting tools' lack of integration with design work (Section \ref{s:formative-lacking-support}). Although LLMs present a new and exciting material for designers, there was no means to entangle the material with their existing processes.

We thus propose a design code, which we call \textit{designerly adaptation}. We define designerly adaptation to be a translational process by which designers \textbf{entangle process with material by translating their requirements into designs for LLM-powered UX---and vice versa---through adapting LLM behavior.} Our proposed code is \textit{designerly} in nature because the resulting designs (the LLM-powered UX) embody design knowledge constructed from intersecting processes with materials. To further characterize designerly adaptation, we present four design goals to guide its practical implementation.

\subsection{Design Goals}
\label{s:design-goals}
Prompting does not serve as a satisfactory design code because 1) it is often divorced from  design practices, environments, and abstractions; and 2) it is unclear how prompts can be used to express design requirements. These realizations informed our design goals below to guide our efforts of operationalizing designerly adaptation with a \replaced{technology probe}{design tool}.
\begin{itemize}
    \item[\textbf{DG1:}] \textbf{Integrate seamlessly with existing design environments and abstractions.}
    \\Designerly adaptation is, at its core, a design practice. Given designers' preference to stay in one tool of their choice (e.g., Figma) for most stages of the design process \cite{feng2023understanding}, we aim for \canvil{} to smoothly integrate with existing design tools and leverage common abstractions within those tools (e.g., layers, components, frames) to align with designers' mental models. Indeed, we saw designers discarding tools like ChatGPT and prompt playgrounds because they did not fit into their design workflows or were overly technical (\ref{s:formative-lacking-support}). 
    
    \item[\textbf{DG2:}] \textbf{Support direct translation of design requirements into model behavior.} 
    \\
    As noted in Sections \ref{s:formative-involvement-normative} and \ref{s:designerly-adaptation-def}, designers lacked the codes to directly translate and apply envisioned constraints in their design work. This undermines designers' contributions to LLM-powered products---after all, they are well-positioned to understand this context using user-centered methods and synthesize their understanding and findings into constraints on model behavior (\ref{s:formative-involvement-normative}). With designerly adaptation and \canvil{}, we provide a new avenue of translation between requirements and LLM-powered UX to empower designers' participation in model adaptation.

    \item[\textbf{DG3:}] \textbf{Enable iterative model tinkering within the design process.}
    \\Designers found model tinkering highly valuable in our formative study as it allowed them to quickly understand LLMs as a design material---how they behave, what they are capable of, and where their limitations are (\ref{s:formative-adaptation-materialistic}), aligning with findings from prior work \cite{subramonyam2021protoai, feng2023addressing}. A design code that encourages tinkering can therefore support rapid ideation, testing, and iteration of many possible solutions and foster productive co-evolution between design requirements and designed objects. 

    \item[\textbf{DG4:}] \textbf{Provide opportunities for collaboration.}
    \\UX practice is highly collaborative \cite{feng2023understanding, deng2023fairness}. Designers told us that they collaborated closely with not only other designers, but also product managers, software engineers, data scientists, and more (\ref{s:formative-adaptation-materialistic}). Designerly adaptation may be primarily performed by designers but should welcome collaboration with non-design stakeholders. This not only allows more diverse insights to be incorporated into adaptation, but also catalyzes further refinement of adapted models (particularly by technical teams) towards production readiness.
\end{itemize}

\section{\canvil{}: a Technology Probe for Designerly Adaptation}
In light of our formative study and design goals, we introduce \canvil{}\footnote{The name ``Canvil'' is a portmanteau of ``canvas'' and ``anvil.'' We envision \canvil{} to be a metaphorical anvil by which LLM behavior can be shaped within design canvases.}, a technology probe for designerly adaptation in the form of a Figma widget. In this section, we describe \canvil{}'s design choices, user interface, and implementation.

\subsection{\canvil{} as a Probe}
\label{s:probe}
We designed \canvil{} as a \textit{technology probe} to investigate whether and how designerly adaptation may be included in existing design workflows. Commonly used in contextual research in HCI \cite{jorke2023pearl, hohman2019gamut}, a technology probe is an artifact, typically in the form of a functional prototype \cite{hutchinson2003technology}, presented to the user ``not to capture what is so much as to inspire what might be'' \cite{lury2012inventive}. That is, probes offer one instantiation of tooling and/or interaction techniques for a domain to better understand phenomena within that domain. Hutchinson et al. \cite{hutchinson2003technology} state that technology probes have three goals: the \textit{social science} goal of understanding users in a real-world context, the \textit{engineering} goal of field-testing the technology, and the \textit{design goal} of inspiring new technologies. We map these three goals onto our objectives with \canvil{}:
\begin{itemize}
    \item \textit{Social science:} understand how engagement with designerly adaptation impacts designers' work on LLM-powered user experiences through a structured design activity in Figma.
    \item \textit{Engineering:} develop and launch a Figma widget that connects to LLMs via APIs and allows them to be adapted and used from within Figma.
    \item \textit{Design:} encourage reflection on designerly adaptation as a UX practice and inform future tools to support it.
\end{itemize}

\added{We note that because probes are investigative tools for studying existing or new phenomena, they are evaluated not through traditional comparisons with baselines, but by their ability to shed new light on the phenomena of study} \cite{hutchinson2003technology, jorke2023pearl, hohman2019gamut}. \added{We demonstrate the effectiveness of \canvil{} as a probe by using it to answer our research questions about designerly adaptation posed in Section} \ref{s:design-study}.

\subsection{User Interface}
\label{s:canvil-ui}
The \canvil{} interface resembles an interactive card. The card itself is separated into two areas: the \textit{Main Form} and \textit{Playground Area}. When a user selects a \canvil{}, a property menu is invoked that can open up additional panels for styling, response generation, and settings. \canvil{}s can be freely placed on and moved around the Figma canvas (\textbf{DG1}), allowing model tinkering to take place in close proximity to relevant designs. \canvil{}s can also interact directly with designs by reading inputs from and writing model outputs to their text layers. We detail each of \canvil{}'s features in this section and connect them with our design goals. 

\begin{figure*}[h]
    \centering
    \includegraphics[width=0.9\textwidth]{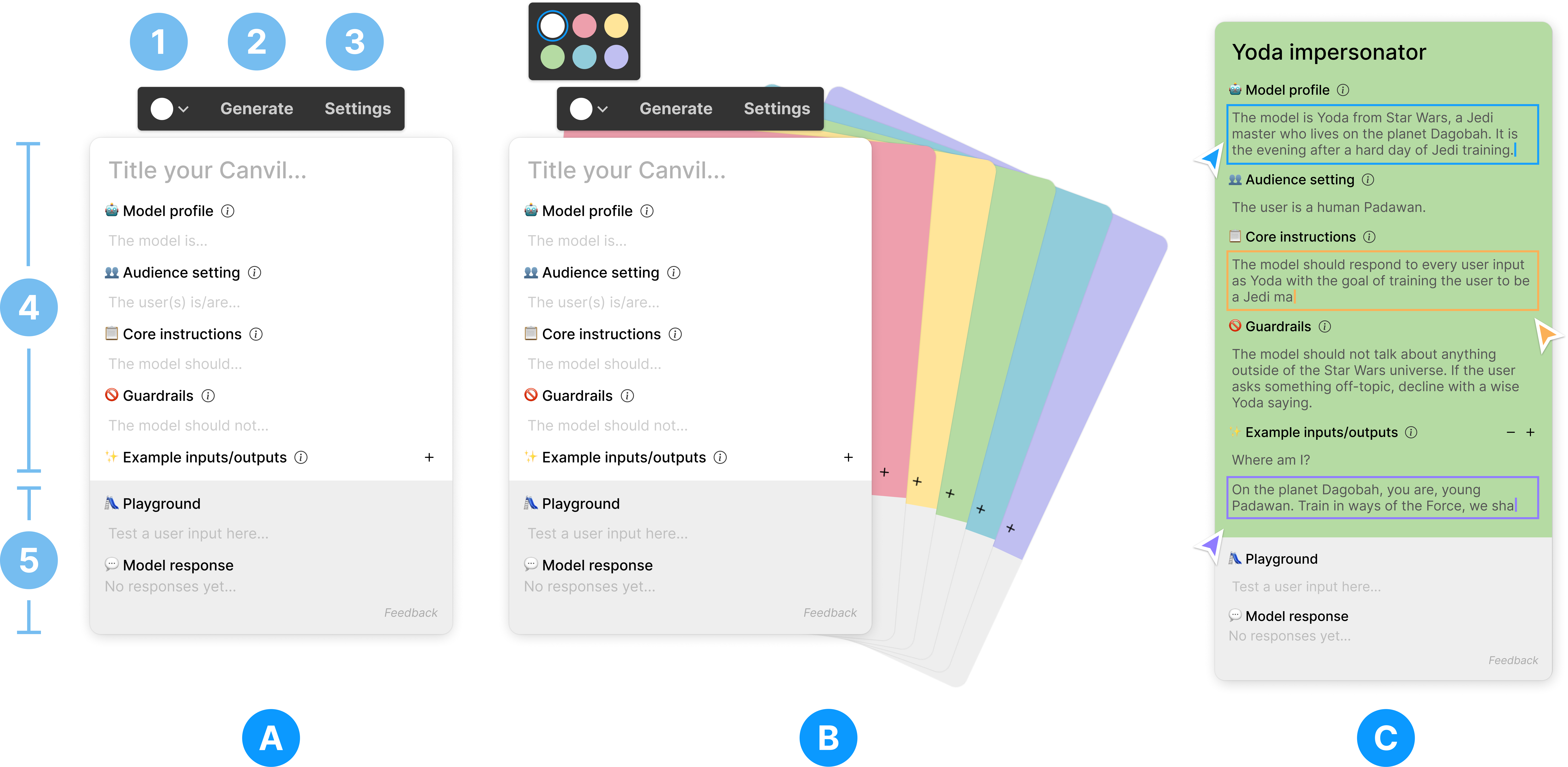}
    \caption{An overview of the \canvil{} interface. \textbf{A:} A blank \canvil{} with its property menu invoked, with some additional copies styled with pre-set color options. The property menu has options for styling \textbf{(1)}, model response generation \textbf{(2)}, and model settings configuration \textbf{(3)}. The \textit{Main Form} \textbf{(4)} contains text input fields for structured authoring of model behavioral specifications. The \textit{Playground Area} \textbf{(5)} allows users to quickly test inputs and view model outputs. \textbf{B:} \canvil{} supports collaboration by default. Here, we see Figma users collaboratively authoring a \canvil{} titled ``Yoda impersonator.'' Like any other native Figma object, \canvil{} also supports stateful duplication.} 
    \label{fig:canvil-overview}
    \Description{On the left, a white card that resembles a form (the Main Form) with a section at the bottom in light gray (the Playground Area). A dark gray property menu is on top of the card, with 3 options: a color picker, Generate, and Settings. On the right, a green card titled Yoda impersonator with text populated in each of its fields is collaborative edited by multiple cursors.}
\end{figure*}

\subsubsection{Main Form} 
The Main Form provides a structured means of authoring model behavioral specifications in natural language via a multi-field form. 
Unlike prior systems that offer an open text field for defining similar specifications \cite{openai-custom-prompts, petridis2023promptinfuser, petridis2023promptinfuser-lbw}, we chose to enforce structure because it provides mental scaffolding to reason about model behavior in a UX context from multiple facets (e.g., \textit{with whom} will the LLM interact, and \textit{how} should the LLM meet their goals?), thus providing more opportunities for designers to translate design requirements into behavioral constraints for the LLM (\textbf{DG2}). Breaking down specifications into smaller units also allows for more fine-grained tinkering and iteration (\textbf{DG3})---designers can copy a \canvil{} and tweak a specific field to compare how that change impacts model behavior. Additionally, because Figma widgets support multi-user interaction by default, designers can collaboratively author specifications (\textbf{DG4}). 

Below are the fields, with a brief description of each, in the Main Form. The field titles, \textbf{in bold}, are always visible on the interface, while the field descriptions can be accessed by hovering over an info icon beside each field title. 
\begin{itemize}
    \item \textbf{Model profile:} High-level description of the model's role, character, and tone.
    \item \textbf{Audience setting:} Persona or descriptions of user(s) who will interact with the model.
    \item \textbf{Core instructions:} Logical steps for the model to follow to accomplish its tasks. Specify input/output format where applicable.
    \item \textbf{Guardrails:} How the model should respond in sensitive or off-topic scenarios, including any content filters.
    \item \textbf{Example inputs/outputs:} Examples to demonstrate the intended model behavior.
\end{itemize}

 We derived these fields from synthesizing best practices from technical tutorials and documentation \cite{openai-system-prompt, msft-system, anthropic-system-prompt} as well as NLP literature \cite{deshpande2023toxicity}. We note that our fields are just one possible way to translate design requirements into model behavioral specifications, and there may be some overlaps between the fields. As discussed in Section \ref{s:rw-llm-adaptation}, no ``gold standard'' currently exists for this translation, so we chose to follow the current recommended best practices when designing the Main Form. 

\subsubsection{Playground Area}
Below the Main Form, we provide a Playground Area as an easy way to send user inputs to the model and test model responses (\textbf{DG3}) within \canvil{}.
The designer may test on the same \canvil{} multiple times or duplicate one with its entire state and test different tweaked versions (Fig. \ref{fig:canvil-overview}B). Designers are likely to find stateful duplication intuitive as it is available on all native objects in the Figma canvas (\textbf{DG1}).

\subsubsection{The Generate Panel}
\label{s:generate-panel}
To generate a response from a model adapted with specifications from the Main Form, the user selects the ``Generate'' option from \canvil{}'s property menu, which takes them to the Generate Panel. The panel has two modes: \textit{Playground} and \textit{Design}. Both modes contain an option to copy the raw prompt to one's clipboard so it can be tested in a separate environment (e.g., a Python notebook during collaboration with a data scientist), if desired. 

\textbf{\textit{Playground Mode}.} The Playground Mode (Fig. \ref{fig:canvil-menus}A) is invoked when the user selects ``Using playground'' from the dropdown on the Generate Panel. This mode instructs \canvil{} to read user input from the Playground Area and write its response back to the Model Response area. This mode is the default mode in \canvil{}.

\textbf{\textit{Design Mode}.} The Design Mode (Fig. \ref{fig:canvil-menus}B) is invoked when selecting ``Using design'' from the dropdown. In this mode, \canvil{} navigates the design layers on the user's Figma canvas, reading text inputs from a specified layer(s) from those designs, and writing model responses to a specified layer(s). This mode leverages the hierarchical layer structure of Figma designs (\textbf{DG1}) and implements the ``input-output'' LLM-interaction proposed by Petridis et al. \cite{petridis2023promptinfuser-lbw}.

\subsubsection{The Settings Panel}
\begin{figure*}[h]
    \centering
    \includegraphics[width=0.85\textwidth]{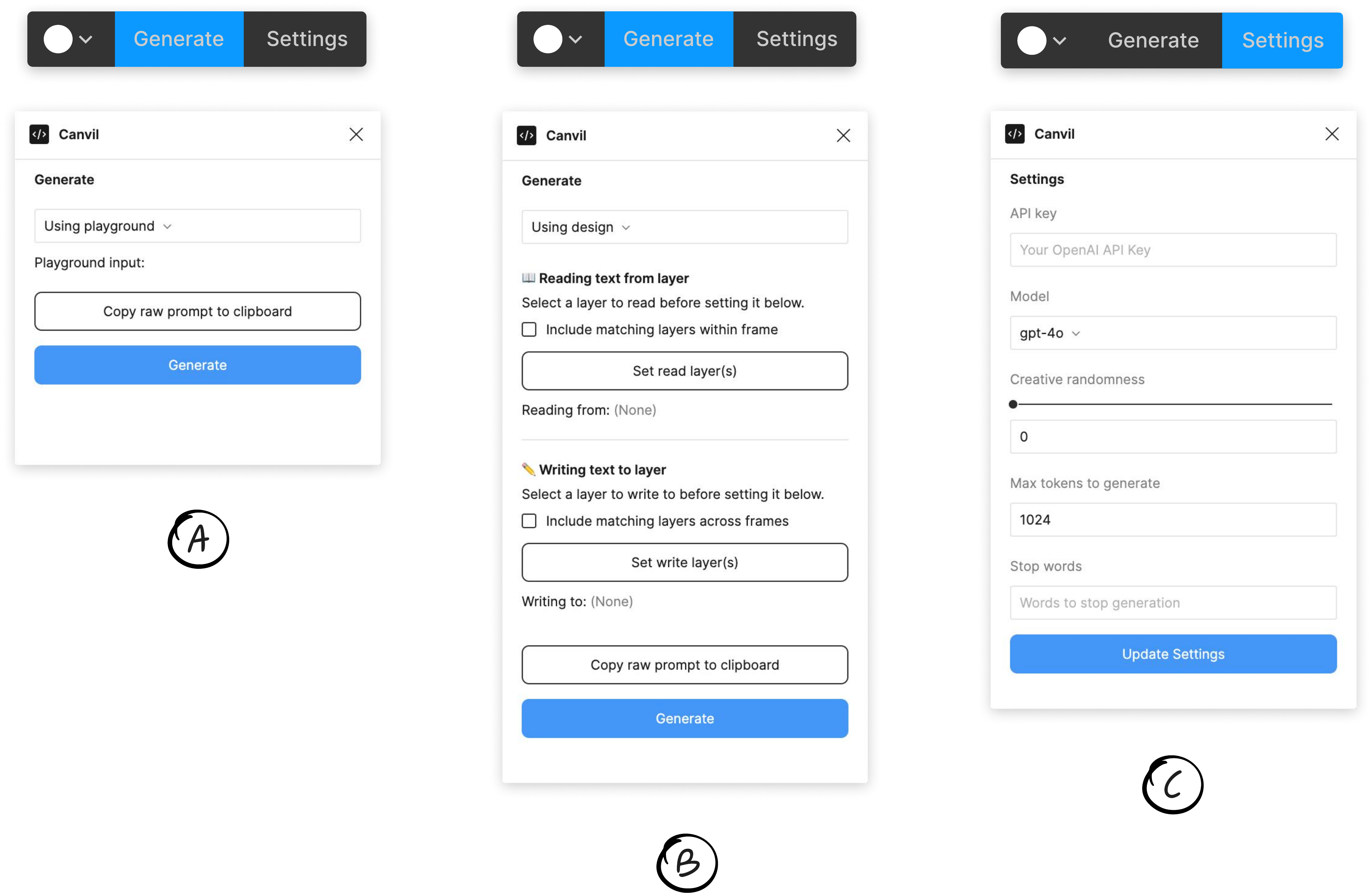}
    \caption{The ``Generate'' and ``Settings'' options in a \canvil{}'s property menu can lead to three panels that appear alongside the \canvil{} itself. \textbf{A:} \textit{Playground Mode} for response generation, where the input is read from and written to the \canvil{}'s \textit{Playground Area}. \textbf{B:} \textit{Design Mode} for response generation, where the input is read from and written to design(s) on the Figma canvas. \textbf{C:} Settings for selecting and configuring the LLM.}
    \label{fig:canvil-menus}
    \Description{From left to right: a panel with an option to generate a model response after reading the playground input; a panel with an option to generate a model response after setting design layers to read the input from and write the output to; a panel with a model selector and model settings.}
\end{figure*}

The Settings panel (Fig. \ref{fig:canvil-menus}C), accessible through the ``Settings'' option from \canvil{}'s property menu, contains some basic model configuration settings. We distilled the list of settings in the OpenAI API to four (model selection, temperature,\footnote{We rename ``temperature'' to ``creative randomness'' to provide a more descriptive name for those not as familiar with LLMs (confusion over the temperature parameter was raised in our formative study---see \ref{s:formative-lacking-support}).} maximum generation length, stop words) that may be useful to non-AI experts such as designers. The ``Update Settings'' button saves the settings to \canvil{}'s state, which is preserved when the \canvil{} is duplicated. \looseness=-1

\subsection{Implementation}
\label{s:canvil-implementation}
\canvil{} is a Figma widget\footnote{https://www.figma.com/widget-docs/} implemented in TypeScript using the Figma API\footnote{https://www.figma.com/developers/api}. A Figma widget differs from a Figma plugin\footnote{https://www.figma.com/plugin-docs/} by its collaborative nature. A widget is available to all users of the canvas that it is placed in, and maintains a common state that supports multi-user editing out-of-the-box. Users can interact with widgets just like any other native object on the Figma canvas, including moving, duplicating, and styling. In contrast, Plugins are local to an individual user. The collaborative affordances of widgets align well with \textbf{DG4}, so we implemented \canvil{} as a widget.

Upon the user selecting ``Generate,'' \canvil{} prepares a system prompt using specifications from the Main Form. Each field in the form is converted to markdown format and is packaged up in ChatLM\footnote{https://learn.microsoft.com/en-us/azure/ai-services/openai/how-to/chatgpt} for added prompt parsability. Our prompt template is available in Appendix \ref{a:prompt-template}. The populated template is sent to an OpenAI Chat API endpoint with settings specified in \canvil{}'s settings panel. If generation was triggered in Playground Mode, any text in the Playground Area will be sent as user input to the model. If generation was triggered in Design Mode, Canvil searches for the read layer(s) specified in Design mode on the user's current Figma canvas, retrieves the text within those layers, and sends them off to the API endpoint as user input. 

\canvil{} is compatible with both the Figma design editor and the FigJam whiteboarding tool. All features have the same behavior in both environments, except that the Design Mode for generation is not available in FigJam because of FigJam's inability to access and edit design elements. 

\section{Design Study}
\label{s:design-study}
 With \canvil{} as a technology probe, we conducted a task-based design study with 17 participants organized into 6 groups. The study was conducted on a per-group basis; each session was 90 minutes in length. Our study investigated the following two research questions:
\begin{itemize}
 
    \item[\textbf{RQ1.}] \textbf{A new design code in practice:} How do designers make use of the two-way translation between design requirements and UX designs enabled by designerly adaptation via \canvil{}, within the context of our study? 
    \item[\textbf{RQ2.}] \textbf{Designerly adaptation at large:} What are designers' attitudes toward designerly adaptation beyond the context of our study? 
\end{itemize}

\subsection{Participants}
We invited participants to our study via email, professional interest groups, Slack channels for designers run by a public U.S. university, and word of mouth at a technology company. Full recruitment details can be found in Appendix \ref{a:recruitment}. 
Invitees first filled out a screening form with information on demographics, professional background, prior exposure to LLMs, and group member preferences. We selected our participants on a first-come first-serve basis, but accounted for their prior experiences designing LLM-powered UX---we aimed for only half our participant pool to have such experience. \added{We did this because we wanted to verify that designerly adaptation was accessible to those working with LLMs for the first time, and that such designers also found \canvil{} to be intuitive}\footnote{In our results, we did not observe noticeable differences in usage patterns the two groups, except that the latter drew more connections between adaptation and their past model tinkering workflows. This was a promising indication that designerly adaptation was accessible to designers regardless of their prior LLM experience.}. We then organized all candidates into groups based on member preferences and availabilities. We left recruitment open as we conducted the studies and stopped when we reached data saturation. In the end, we had 6 groups with 17 participants total; five groups were of size 3 and one of size 2. All participants were based in the United States. Participant and group details can be found in Table \ref{t:participants-2} in Appendix \ref{a:design}. \looseness=-1

All study sessions were recorded and transcribed. We gifted each participant a \$75 USD gift card after the study. This study was approved by the IRBs of all organizations involved.

\subsection{Study Design}
Our task-based study design was motivated by contextual inquiry \cite{beyer1999contextual} due to its ability to elicit rich information about participants' work practices and processes. We conducted our study virtually over videoconferencing software and Figma. We opted for a group study design to better emulate the team-based environment in which many designers now work \cite{feng2023understanding}\replaced{---because collaboration is such an integral part of the design process, we wanted to ensure that designerly adaptation did not only embrace collaboration, but also enrich it. Specifically, we sought to emulate the collaboration modes designers were used to. Prior work found that designers tended to execute on design tasks (wireframing, prototyping) independently but come together to discuss justifications of design choices and create documentation \protect{\cite{feng2023understanding}}. We thus structured the study to have an independent work period followed by synchronous sharing and discussion of designed artifacts (including participants' \canvil{}s). This respects existing workflows while surfacing deeper collaborative insights than prior 1:1 studies}{A group study may also help uncover more insights about enabling collaboration (\textbf{DG4}) than prior individual design studies about AI-powered user experiences} \cite{petridis2023promptinfuser, subramonyam2021protoai, feng2023addressing, feng2023ux}. 

\subsubsection{Setup}
The following instructions for the study's design task were provided to all groups: 
\begin{quote}
    \textit{You are all designers on a product called Feasto. Feasto aims to increase users’ enjoyment of food. The Feasto team is scoping out a new feature called the 3-course meal planner, which allows users to get suggestions for 3-course menus they can cook by simply listing out some ingredients they have. The team agrees that large language models (LLMs) are a promising technology to power this new feature. Your task is to design the UX for the area in which users will see and interact with the suggested menus.}
\end{quote}

\begin{figure*}[h]
    \centering
    \includegraphics[width=1\textwidth]{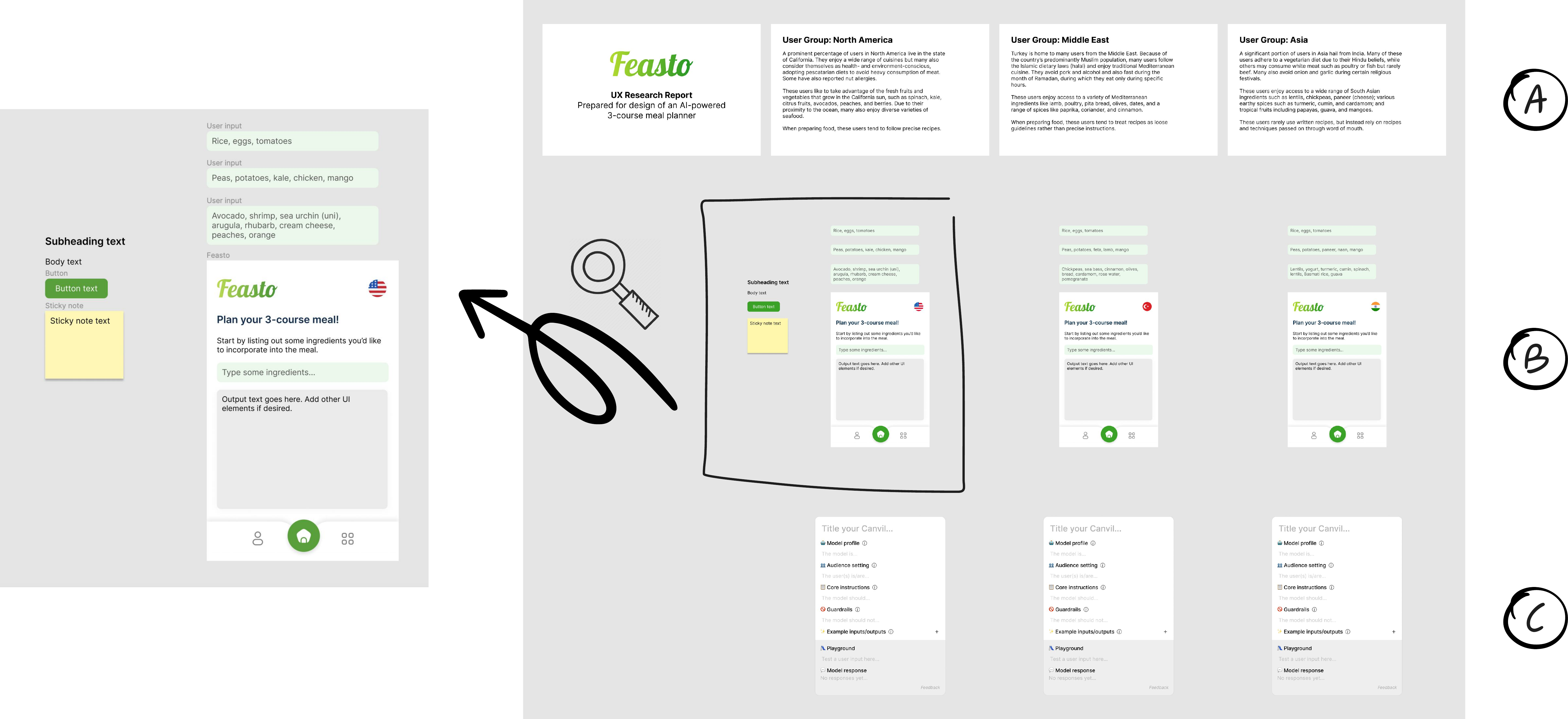}
    \caption{The setup for an individual participant's canvas in our study's Figma file. \textbf{A:} Informational packet containing descriptions of three user groups residing in North America (primarily west coast of the U.S.), the Middle East (primarily Turkey), and Asia (primarily India), respectively. \textbf{B:} Starter UIs for Feasto's 3-course meal planner with example user inputs to lower the barrier for testing model responses, along with basic UI elements such as text and buttons. \textbf{C:} Blank \canvil{}s for participants to adapt LLMs.}
    \label{fig:board-setup}
    \Description{A Figma canvas with three sections. The top section has a 4-page document containing text descriptions of three user groups for participants to work with during the study. The middle section has starter UIs and UI elements for Feasto's 3-course meal planner feature. The bottom section has three blank \canvil{}s for participants to adapt LLMs.}
\end{figure*}

As part of the exercise, we intentionally left designers to determine how exactly the LLM powers this feature. We also assembled these instructions cognizant of the fact that this was a time-constrained study, and we wanted participants to specifically focus on the area of the interface where users interact with an LLM. To reduce design overhead unrelated to our research questions, we provided starter UIs for participants to work with, along with basic UI components such as text, buttons, and sticky notes (Fig. \ref{fig:board-setup}B). If participants had an idea but did not have time to execute it, we encouraged them to describe it on a sticky note next to their designs. Finally, we picked a universally understood domain (food and meal planning) to avoid any variance in domain-specific knowledge.

To better probe \textit{research-informed} model adaptation---that is, adaptation across varying user contexts informed by user research (\textbf{DG3})---
we created descriptions for three hypothetical user groups of this new Feasto feature based on users' geographic region: west coast of the U.S., Turkey, and India. These descriptions are not meant to act as user personas, but rather high-level sketches of the customs and preferences that may be prevalent in the region. We crafted the user group descriptions to vary along three key dimensions (assuming everyone used Feasto in English): \textbf{dietary restrictions}, \textbf{access to ingredients}, and \textbf{menu style preferences}. Definitions for these dimensions, along with the full user group descriptions, are available in Appendix \ref{a:user-groups}. 

We provided one blank starter UI with some example user inputs and one blank \canvil{} per user group. Participants were invited to vary the user experience between user groups as much or as little as they saw fit. We also encouraged participants to use \canvil{} to adapt an LLM and test its behavior as they designed. 

A Figma file for each study group contained all the materials described above. Within the file, we created separate canvases for each participant to act as individual workspaces, each with its own copy of the materials (Fig. \ref{fig:board-setup}). We also had a shared canvas for introductions and instructions before the task and collectively debriefing afterwards. 

\subsubsection{Procedure}
Our 90-minute study was divided as follows.

\textbf{Introduction (20 minutes):} First, all participants introduced themselves to others in the group. The study facilitator then gave a demo of \canvil{}, covering all features described in Section \ref{s:canvil-ui} using a pre-filled \canvil{}. The facilitator also described the design task and answered any clarifying questions from participants.

\textbf{Design task (40 minutes):} Participants spent 40 minutes on the design task and were asked to consider at least two of the three user groups provided. Participants were encouraged to spend 10 minutes authoring a \canvil{} and another 10 minutes on UI design per user group. Some participants who had remaining time designed for all three user groups.

\textbf{Group interview (30 minutes):} Participants first filled out a brief usability questionnaire about \canvil{} before gathering in the shared page of the Figma file. They were asked to copy their \canvil{}s into the shared space and also their designs (if desired) to share with the group. The facilitator then led a semi-structured interview that asked participants to reflect on their experience adapting models with \canvil{}, \canvil{}'s collaborative capabilities, and how they see adaptation fitting in with their own design practice. The facilitator ensured that each participant had ample opportunity to express their thoughts in the group setting, and also encouraged dialogue between participants.

\subsection{Data Analysis}
We conducted a qualitative analysis of transcriptions and Figma canvases (including authored \canvil{}s), as well as a quantitative analysis of feedback from \canvil{}'s usability questionnaire.

For our qualitative analysis, the first author took a hybrid inductive-deductive approach to coding the group interview portion of the transcriptions from the study \added{in the Marvin qualitative coding tool.} This process started with an open coding round in which high-level themes were generated, followed by subsequent rounds of thematic analysis via affinity diagramming in which themes were broken down into sub-themes. This approach was taken because new subtleties and complexities emerged from our initial codes as coding progressed due to the diverse approaches observed in our study as well as participants' group discussion dynamics. The codes and themes were discussed and iterated on with research team members at weekly meetings. Additionally, whenever participants made references to content within their Figma canvases (e.g., their designs and/or \canvil{}s), the first author took screenshots of those references and linked them to transcript dialogue. Summary memos were then written for our high-level codes and presented alongside relevant screenshots.

\canvil{}'s usability questionnaire followed the standard template for the System Usability Scale (SUS) \cite{sus-scale, brooke1995sus} and consisted of 10 questions, each with five response options for respondents from Strongly agree to Strongly disagree. We computed a SUS score using methods outlined by Brooke \cite{brooke1995sus} for each participant and subsequently computed a mean score and standard deviation.

\section{Results}

In this section, we \replaced{demonstrate the effectiveness of \canvil{} as a probe by sharing the takeaways from our two research questions related to designerly adaptation through qualitative insights from our design study.}{focus on the qualitative insights from our design study that shed light on our two research questions.} We present text that participants wrote in \canvil{} as \textcanvil{purple and italicized}. As a basic usability check, we conducted quantitative analysis to show that \canvil{} had ``above average'' usability, with a mean SUS score of 69.94 (std = 12.18) \cite{sus-scale}. 

\subsection{RQ1: A New Design Code in Practice}
\subsubsection{From Design Requirements to UX Designs} 
\label{s:findings-req2design}
\begin{figure*}[h]
    \centering
    \includegraphics[width=1\textwidth]{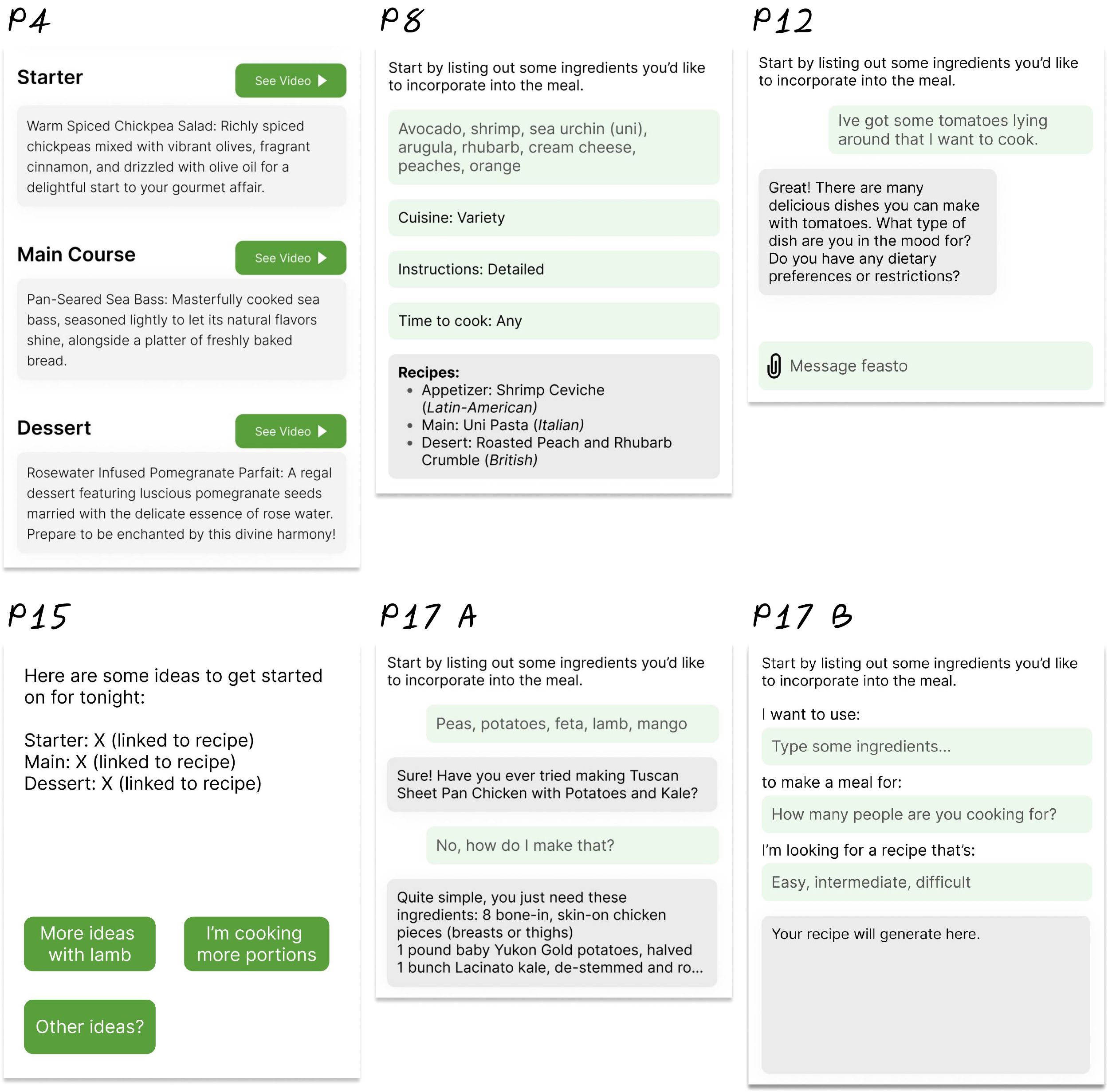}
    \caption{Examples of UI screens designers created for Feasto's 3-course meal planner during the study.}
    \label{fig:ui-showcase}
    \Description{6 UIs showing different approaches designers took to designing the UI for Feasto's 3-course meal planner, including simply displaying the output, chat interfaces, and additional buttons.}
\end{figure*}

Designerly adaptation, as a design code, is meant to facilitate translation of design requirements to concrete designed objects---in our case, UX designs. We saw robust evidence of designers effectively leveraging this translation in our study.  \canvil{}'s integration into the Figma design environment (\textbf{DG1}) and its support for quick iteration (\textbf{DG3}) allowed designers to more deeply reason about user-LLM interactions based on model behavior shaped by their design requirements. 
P17, for example, tried two different interaction patterns for their interface---a chat-based interface versus a form-filling GUI---to see which one better handles LLMs' high sensitivity to their specifications' wording. P15 provided alternatives via buttons to support more flexible user input of ingredients to the model, while P8 described their UI design in response to observing the limits of what can be achieved via editing the Main Form: \textit{``Sometimes even in the instructions that I gave to \canvil{}, it wasn't really reflecting that [desired] granularity until I pushed it further. In my design I ended up putting a little textbox area where people can specify how detailed they want the instructions.''} Examples of quoted participants' designs can be found in Fig. \ref{fig:ui-showcase}. 

Interestingly, a few designers wanted a degree of separation between UI work and model tinkering. P1 shared that they found it \textit{``a bit hard to juggle between \canvil{} and [the UI] at the same time''}, especially when the output is incompatible with their design settings: \textit{``What if the text is really long and then I have to play with auto layout?''} Therefore, P1 preferred first tinkering with the model using the Playground Mode. P8 also agreed that having \canvil{} inject outputs directly into text boxes can be \textit{``a little scary [...] people don't wanna actually commit [responses] to text boxes sometimes.''} Past work considers integration between models and UI designs as desirable \cite{petridis2023promptinfuser, petridis2023promptinfuser-lbw, subramonyam2021protoai}. Our results encourage providing designers with more choices and control over such integration.

\subsubsection{From UX Designs to Design Requirements} 
\label{s:findings-design2req}
Design codes facilitate translation from not just design requirements to designed objects, but also from objects back to requirements \cite{cross1982designerly}. Here, we discuss the reverse translation of Section \ref{s:findings-req2design}. Some designers were more adamant about adapting the model to fit existing UI. P4 mentioned that they \textit{``iterated multiple times''} on the Main Form to find the ideal response. They often stopped iterating when they viewed that the LLM output satisfied the UX requirement, such as being concise: \textit{``what I like the best about this [response] was that it was short and [ideal] for scrolling in a screen. I tried that out [in my new UI] and it worked pretty well''} (see Fig. \ref{fig:ui-showcase}). 

Many designers also aimed to maintain a consistent UI across the user groups they were designing for, but adjusted the requirements they typed into \canvil{} to create meaningful customizations of LLM based on nuanced differences in user needs. This was enabled by \canvil{}'s main form, which implements \textbf{DG2}. For example, in the \canvil{} for the user group from India, P4 wrote that \textcanvil{The model should use Indian terminologies to describe the recipe. Eg: eggplant is brinjal in India.} P5 experimented with different \canvil{} versions \textit{within one user group} and assigned distinctly different personalities for each model. For their user group in India, they populated one \canvil{}'s Model Profile field with \textcanvil{The model is a head chef of a 5 star restaurant situated in New Delhi. It is a busy day for the restaurant and the chef is low on time.} and another with \textcanvil{The model is a mother who is helping her son cook quick meals in hostel.} After some experimentation, they noted that \textit{``the models did very well''} in taking on these different profiles. 

Iterating on the Guardrails field in \canvil{} specifically allowed designers to address potential LLM safety concerns, such as violations of users' dietary restrictions and model misuse. In addition to stating dietary restrictions of their Turkish user group in the Guardrails field, P12 also specified what the model should do in response: \textcanvil{The model should not include recipes with pork or alcohol and must respect the fasting period of Ramadan by suggesting suitable pre-dawn and post-dusk meals.} A few designers noted that some user groups may require more strict guardrails than others. P15 identified a subtle but key requirement between the \canvil{}s for the American and Turkish user groups:

\begin{quote}
    \textit{``I wanted to highlight that the model could take many more liberties with the recipes that it was giving [to the Turkish user group], but it couldn't take more liberties with the ingredients. It could be very loose with: try this, try this with this, but never like crossing the boundary of the dietary restrictions. Whereas with the [American] one, it'll take it into account, but it's not gonna mar their religious practice.''} [P15]
\end{quote}

UX design was never just about the visual UI---user flows, content strategy, customization heuristics, and many non-UI considerations play vital roles in shaping a user experience. We can see through our findings that this is especially true with LLM-powered UX. Even within the same UI, designers may create entirely different experiences based on their (shifting) design requirements by adapting LLMs via \canvil{}. Thus, through designerly adaptation, designers can more effectively explore and use LLMs as a design material.

\subsection{RQ2: Designerly Adaptation at Large}

\subsubsection{Designers Were Receptive to Designerly Adaptation but also Recognized its Limits}
\label{s:process-integration}

Designers saw great value in engaging with adaptation and saw direct paths to application in their own design practice beyond the study. P10, who has extensive experience working on AI-powered products deployed to users worldwide, said that working with AI as a design material has traditionally been difficult: \textit{``We had teams of people training models and nudging the technology to align with [user personas].''} Having experienced LLM adaptation via \canvil{}, they shared that \textit{``I can see this being something that would enable [adaptation] to actually happen in a way that's much more practical than taking Python classes, which I've done.''} For P13, who works on a product with enterprise and consumer versions, adaptation can help set more detailed product requirements where necessary: \textit{``I could see [adaptation] being really helpful, being able to prototype and build separate generative models for enterprise customers and then to work more closely with them to tune and add requirements.''} P16 considered adaptation to be a useful exercise for user empathy: \textit{``for [Audience Setting] I just tried to put myself in the mind of someone who's using this and say, I'm really good at following recipes, but have a variety of dietary preferences and restrictions.''}

Designers also recognized that designerly adaptation is only one piece of the larger puzzle when it comes to defining an LLM's behavior in a production-ready product. Some changes desired by designers, such as factual grounding and connecting to external knowledge bases, may still need engineered by technical teams. Both P12 and P14, for example, wanted \canvil{} to be able to connect to custom models built and maintained by their own teams, as the behavior of default OpenAI models may be too generic to use as a suitable design material. P14 comments: \textit{``I think linking to the vanilla ChatGPT model isn't specific enough. If there is a way to somehow have \canvil{} link to the devs' [models] while I'm designing and I'm able to see what type of experience our users are getting, I think that would be super helpful.''} While we currently offer only a limited number OpenAI models to use with \canvil{}, custom models may be added with lightweight engineering as the widget is communicating with the model via API endpoints. This would also \canvil{} to integrate technical advancements into design processes. 

\subsubsection{\replaced{Designers Identified Promising Collaborative Potential for Designerly Adaptation}{\canvil{} Afforded Collaboration and Knowledge Sharing Within the Design Process}}
\label{s:collab-knowledge-sharing}
\replaced{Because of \canvil{}'s seamless integration with Figma and its collaborative canvases, designers envisioned engaging with designerly adaptation in collaborative ways (\textbf{DG4}) by using existing, already-familiar multiplayer affordances. For example, P5 appreciated being able to overlay comments on top of \canvil{}s: }{Many designers commented on how \canvil{} afforded collaboration (\textbf{DG4}), a much-welcomed benefit that extends beyond the context of our study. P5 appreciated that they could leverage Figma's built-in collaboration features:} \textit{``If I see my fellow designer's \canvil{}, and if I want to change something then and there, I can drop a comment on it just like a normal component in Figma, which we've been doing in our everyday design work.''} \deleted{P10 shared that their team relies on a spreadsheet to keep track of model adaptation efforts and working in a canvas environment would be a significant improvement.}

In particular, designers saw potential for \canvil{} to facilitate more effective knowledge sharing about LLM adaptation across design and non-design teams. P10 imagined that: \textit{``You could have a master \canvil{} and then you could make copies, and [others] can then do their own interpretations.''} P11 agreed and added that seeing iterations on a canvas can help find inspiration in others' work: \textit{``I think anytime you line up different iterations together, you notice: ohh that person, did you know they had that approach? That's a good idea. And I'm gonna try that over here. I think it really is an aid to experimentation.''} 

We observed other instances of knowledge sharing when designers viewed others' \canvil{}s. For example, P4 noted that they were inclined to consider more dimensions for their Main Form after seeing P6's work: \textit{``[looking at P6's \canvil{}] made me thinking about timers and Hindi slang.''} Some also realized new capabilities of LLMs by looking at others' Main Form and outputs---P1 shared that \textit{``I didn't realize at first that you can actually make the [LLM] generate multiple recipes, just like what P2 did.''} P1's groupmates (P2 and P3) were also intrigued when they saw that P1 had used one \canvil{} to generate example inputs for another \canvil{} to use. These social and collaborative affordances differentiate \canvil{} from prior systems for empowering designers to tinker with AI \cite{subramonyam2021protoai, petridis2023promptinfuser, carney2020teachable}. \replaced{More broadly, \canvil{} can serve}{Designers were also excited by the use of \canvil{}} as a boundary object \cite{star1989institutional} to collaborate on model adaptation \textit{across domain boundaries}. P2 shared that they \textit{``can see from a product manager's perspective that they would love to play around with [\canvil{}], and it would probably help the designer find more common ground because they both used a similar tool.''} For deepening collaboration with data scientists and other stakeholders, P11 commented that the ``copy to clipboard'' feature in Generate Panel (Section \ref{s:generate-panel}) can allow for direct handoff of model behavioral specifications: \textit{``[The button] is a way of exporting that so a data scientist can come along and say sure, let me plug that into code.''} Even if designers were unable to achieve the desired model behavior themselves, P17 thought \canvil{} was helpful in specifying desired changes: \textit{``If I want the model to respond in this way versus that way, just having something tangible to show engineering partners where the tweak would need to be, would be helpful. And I think [that's] obviously easier with \canvil{}.''}

\subsubsection{Desired Areas of Additional Support}

Designers identified several areas of improvement to better support designerly adaptation. For one, designers lacked a clear mental model of how the fields of \canvil{} would impact the adaptation outcome, including how much detail is required in authoring the specifications. As a result, some designers wrote long, detailed instructions, while others kept their \canvil{}s sparse, and some were pleasantly surprised by how well the model handled minimal instructions. P6 further noted that the stochastic nature of the model was particularly challenging to work with when writing outputs to their designs: \textit{``every time I generated the [response], I felt that they were different every time, so it was not easy to predict what the next [response] would look like.''} We note that current LLMs are known to be challenging to control precisely through natural language, and there is a lack of transparency into (or even an established understanding of) how natural language instructions influence LLM behaviors. However, this is an area of active research which can help improve \canvil{} users' mental models and the tool's general utility. For example, new developments in using sparse autoencoders to surface human-interpretable features in LLMs as novel controls \cite{templeton2024scaling, gao2024scaling} may be integrated into \canvil{} to help designers steer models using specific features relevant to their design problem. 

Additionally, some designers wanted to engage in finer-grained experimentation and iteration by only focusing on a specific field in the Main Form. P7, who wanted to iterate more on the Model Profile field, wondered if there was a way to \textit{``decrease the size [of the other fields] and expand to view everything.''} We envision a modular future version of \canvil{} where each field can be separated, such that a user can mix-and-match different fields. However, before that, we may need to address the precise mapping between each field (and potential overlaps between them, as noted by some participants) to adaptation outcomes as discussed above. As mentioned in Section \ref{s:canvil-ui}, our fields present just one possible structure for model behavioral specifications via design requirements. The fields can perhaps be reconfigured to reduce potential overlap, or even dynamically generated based on the design task. 

On a higher level, it would be irresponsible to assume that designers can walk away with a comprehensive understanding of model behavior after a few rounds of tinkering in \canvil{}. While observing a few informative output instances of model behavior aids the design process, formal evaluations ensuring comprehensive coverage of the user input space are crucial for production-ready systems. Thus, new evaluation tooling and processes that loop in technical stakeholders may be required.

\section{Discussion}
\deleted{Our findings shed light on possible workflows for designerly adaptation, promises of tools for collaborative AI tinkering, and implications of materiality on the social and collaborative practices of product teams. We discuss each below.}

\subsection{A Workflow for Designerly Adaptation}
\label{s:workflow-def}

Drawing on the results from both our formative and design studies, we now propose a workflow for designerly adaptation to concretely illustrate how this practice might be used as a design code. \added{This workflow draws heavily from---and intentionally aligns with---those in human-centered design so that designers can easily adopt it when working on LLM-powered applications.} We intentionally make this workflow agnostic to the specific LLM adaptation technique---if a new technique provides finer control over model behavior and is accessible to generalist audiences (e.g., feature clamping \cite{templeton2024scaling}), the workflow can incorporate that technique. Our proposed workflow consists of four steps: 
\begin{enumerate}
    \item \textbf{Understand deployment context through user research.} To orient adaptation and establish user requirements, it is imperative to first understand the context in which users will interact with the LLM-powered application. This includes users' goals, needs, and pain points, along with customs and values that may affect their use of the technology. This step should ideally be led by those with expertise in user research methodologies.
    \item \textbf{Translate user requirements into model behavior, learning from examples where possible.} Appropriate tooling can carve out a direct path for user research to define design requirements that impact model behavior, as we observed in Section \ref{s:findings-design2req}. Thanks to the collaborative nature of many modern design tools \cite{feng2023understanding}, there may be example adaptation attempts by other designers available for reference, or templates to use as a starting point (Section \ref{s:collab-knowledge-sharing}). By leveraging collaborative affordances to share knowledge, designers can more quickly familiarize themselves with the new design material.
    \item \textbf{Co-evolve designs and model behavior.} As observed in Section \ref{s:findings-req2design}, model adaptation can supply new inspiration for UX designs and UI affordances. On the other hand, designers also tinkered with the Main Form to steer the model towards providing outputs that fit into the constraints laid out by existing designs. We see the co-evolution of designs and prompts as a promising path forward, in which iterative tinkering with adaptation approaches shapes design decisions, and vice versa.
    \item \textbf{Share adaptation efforts with the broader team.} Showcasing in-progress work through design critiques is already a part of the design process \cite{feng2023ux}. In our study, we found that sharing \canvil{}s helped envision new collaborative workflows with other designers, as well as communicating their perspectives and negotiating with technical stakeholders (Section \ref{s:collab-knowledge-sharing}). We thus believe that creating shareable artifacts that depict adaptation efforts---i.e., translational knowledge from the use of a design code---is an integral part of designerly adaptation.
\end{enumerate}

Our proposal is not meant to constrain what workflows for designerly adaptation can possibly look like, but rather to offer a concrete entry point for practitioners and researchers to further explore and iterate on this new practice. We invite the community to experiment with this workflow in future practice and research.

\subsection{Navigating Tradeoffs Between Model Adaptation Techniques}

\added{When developing LLM-powered applications, individuals or organizations face several options for tailoring a model to their specific context. These include training their own LLM, adapting an existing LLM through fine-tuning, or using various prompting techniques. Each option differs in complexity and the technical or domain expertise required, and involves trade-offs between robustness, speed, and cost. Many individuals or organizations may only be able to pursue lower-cost, lightweight adaptation---the type afforded by \canvil{}. Lightweight adaptation also happens to be more accessible for designers, which was a major reason we pursued it in our work. While further research is needed to determine the differences in outcomes between lightweight prompting and other adaptation methods for customizing LLMs, it is important to note that in some cases, designerly adaptation may not produce a production-level model.}

\added{In such cases, it may be helpful to consider the analogy of low and high fidelity prototypes, which have similar tradeoffs. A low-fidelity prototype can be created with limited effort and thus allows for quick iteration and divergent exploration of possible alternatives, but sacrifices functionality and faithfulness towards the eventual interface. A high-fidelity prototype is a more faithful representation, but it is also more costly to create and less amenable to iteration. Designerly adaptation can be considered a lightweight, low-fidelity approach to adapting models. It is particularly well-suited for the design process because it can easily support rapid experimentation and iteration. Custom model training and fine-tuning, on the other hand, is high-fidelity adaptation. It improves the robustness of model behavior such that the model can be reliably used in production, but is a costly process often involving bespoke data collection, annotation, and cleaning \cite{ouyang2022rlhf, dodge2020fine, wei2021finetuned}. Because custom training is costly to iterate upon, it can often be practical to first use low-fidelity approaches, like designerly adaptation, to ensure that training efforts are aligned with desired outcomes.}

\subsection{Towards Design Codes for Collaborative Material Exploration of AI}
\label{s:collab-tinkering}

Our work highlights the promises of not only empowering designers to explore LLMs as a novel design material, but also \textit{collaboratively} doing so. For example, designers thought they benefited from seeing and learning from others' \canvil{}s and commented on more organized knowledge sharing and version management afforded by using \canvil{} in Figma's multiplayer canvas (Section \ref{s:collab-knowledge-sharing}). While many tools from prior work lower the barrier for exploring AI's---not just LLMs'---material properties (e.g., \cite{lobe, liner, carney2020teachable, subramonyam2021protoai}), few offer collaborative affordances. To extend existing tools along a collaborative dimension, we encourage a shift from solely focusing on new tools to \textit{design codes which a tool can help operationalize}. 

Because design codes entangle underlying work processes with material exploration \cite{cross1982designerly}, they urge tool builders to consider how tools enabling this exploration can integrate with existing collaborative workflows. Thus, in our work, we developed \canvil{} as a Figma widget to seamlessly integrate into Figma canvases where designers already have well-established collaborative practices \cite{feng2023understanding}. 
Developing a tool \replaced{for individualized workflows}{designing for collaborative workflows early on,} and only later highlighting its collaborative potential may not be sufficient to make it truly collaborative.
For example, node-based editors for steering LLMs (e.g., \cite{Arawjo_2023, suh2023sensecape, wu2022ai, angert2023spellburst}) are conceptually appealing for multiplayer collaboration \cite{wu2022ai}, but few support it in practice. 
\replaced{Extending these tools along a collaborative dimension not only improves their integration into existing collaborative processes, but also enables \textit{collective sensemaking and learning} to accelerate practitioners' understanding of a new design material \protect{\cite{liao2023designerly}}. Applying this to node-based editors, one may consider how a collaborative ChainForge \protect{\cite{Arawjo_2023}} can foster peer learning in prompt engineering, or how a collaborative Spellburst \protect{\cite{angert2023spellburst}} can leverage and extend existing community showcases for creative coding \protect{\cite{p5-gallery}}.}{When embedding collaboration into these tools, one may consider how the tool can entangle existing collaborative processes---which may vary significantly from community to community---with LLMs. For example, understanding how development teams share prompts and model outputs may provide valuable insights for a more collaborative ChainForge \protect{\cite{Arawjo_2023}}, while examining social practices of the p5 creative coding community can inform collaborative features for Spellburst \protect{\cite{angert2023spellburst}}.}

We note that robust collaborative experiences, particularly real-time ones, can be challenging to implement. We tackled this in our work by building on top of the Figma API, which allows us to leverage Figma's built-in collaborative features by default. We thus encourage researchers and practitioners who wish to build collaborative tools to take advantage of existing collaborative platforms' developer APIs where possible, especially as these APIs become more richly featured.

\subsection{Materiality and Sociomateriality of LLMs}

While Cross, in his original 1982 essay \cite{cross1982designerly}, discussed design codes in the context of design education, these codes have notable \textit{sociomaterial} properties when used in collaborative and organizational settings. Designerly adaptation is no exception.
Scholars use the term \textit{sociomateriality} in recognition of materiality's tendency to shape, and be shaped by, organizational practices typically constituted as ``social'' (e.g., decision-making, strategy formulation) \cite{orlikowski2007sociomaterial, leonardi2012materiality}. For example, Orlikowski observed that the issuance of BlackBerry devices within a company led employees to obsessively check for new messages and send immediate responses \cite{orlikowski2007sociomaterial}. The BlackBerry's material properties---in this case, being able to receive and send messages on-the-go---reconfigured employees' social practices, which in turn shifted how they think and act with the technology. 

In our design study, designerly adaptation allowed for interactive exploration of LLMs' material properties. 
Grappling with these properties not only allowed designers to reason about UX improvements (Section \ref{s:findings-req2design}), but also \replaced{serves to better educate designers about LLMs' materiality. This improved ``designerly understanding'' \cite{liao2023designerly} of a new material has the potential to}{demonstrated potential to} reconfigure social practices within product teams. For example, \added{upon tinkering with models and getting acquainted with their materiality,} designers informed us that they saw new avenues of collaboration with non-design stakeholders, which included working with PMs to inform prompts with design requirements, and having a more concrete artifact to communicate desired model changes to engineers and data scientists (Section \ref{s:collab-knowledge-sharing}). Moreover, designerly adaptation may only be one piece of the broader puzzle when it comes to transforming LLMs' materiality at a more fundamental level. Certain capabilities such as storing external knowledge in model weights through fine-tuning cannot easily be unlocked through---and may even a prerequisite for---efforts of teams working at the model level. 
This potential reconfiguration of desigers' collaborative when working with LLMs establishes designerly adaptation as not only a design code, but also a sociomaterial practice. 

The implications of this observation are twofold. First, sociomateriality argues that reconfigurations of collaborative practices upon interaction with a prominent new technology are \textit{inevitable} \cite{orlikowski2007sociomaterial, leonardi2012materiality}. At the time of writing, frenzied excitement over LLM capabilities has launched an industry-wide race to embed them into products and product suites \cite{parnin2023building}. There is much yet to be discovered about shifts in organizational practices and the emergence of new ones in the midst of this race, so researchers in CSCW and organizational science should be attuned to emergent challenges. Second, studying and addressing these challenges may require new processes and tools \added{for educating designers about the materiality of new AI models}. \canvil{} is an early example of such a tool, but more are needed to tackle the multiplicity of open questions in a rapidly evolving AI landscape.

\section{Limitations and Future Work}
Our design study, conducted in Figma, aimed to mirror real-world design activities \cite{double-diamond}, but some concerns about ecological validity remain. User research in practice may differ in presentation and detail than in our study setup, and may not include the use of personas, which can change how designers synthesize user requirements into \canvil{}. Feasto, the fictitious app in our design study, applied LLMs to the universally relatable topic of food. Designers in domains with fewer broadly-shared experiences (e.g., accessibility), may require deeper collaboration with domain experts and thus face workflow complexities not accounted for in our study. A potential direction for future work, then, is longitudinal studies that observe product teams throughout a full development lifecycle of an LLM-powered feature to better understand key adaptation and bolster ecological validity.

Our study, like any study that uses a probe, has results contingent on our probe's features. For example, designers' interaction with the Main Form---designed with recommended practices for defining model behavior using natural language \cite{msft-system, openai-system-prompt, anthropic-system-prompt}---shaped their approach to adaptation, and changes to the form could impact existing results and reveal new ones. We mitigate this by presenting findings not tied to the Main Form, nor our design task. Future research can experiment with different forms for designerly adaptation tooling, or use \canvil{} over longer time periods to surface additional design considerations. 

Designers in our study also provided feedback on \canvil{} that we can integrate into future work. These include breaking down the Main Form into sub-\canvil{}s and linking them together to assemble model behavioral specifications, image generation with multimodal models, and text formatting controls for model outputs. Finally, as mentioned in Section \ref{s:canvil-implementation}, \canvil{} may also be used in FigJam. Future research could explore adaptation in early design stages, like brainstorming and ideation, through studies in FigJam.

\section{Conclusion}
As LLMs become increasingly embedded in our everyday applications, designers are empowered to craft effective LLM-powered UX. Through interviews with 12 designers, we identified a need for a process that can facilitate a two-way translation between LLM behavior and design requirements. We proposed \textit{designerly adaptation} as such a process. We then developed \canvil{}, a Figma widget that operationalizes designerly adaptation by enabling designers to iteratively author, tinker with, and share adapted LLMs as a novel design material within Figma's collaborative canvas environment. We used \canvil{} as a technology probe to \replaced{investigate}{explore} the integration of designerly adaptation in UX practice through a group-based design study with 17 designers in 6 groups. 
\added{Through \canvil{}, we acquired valuable insights into how designerly adaptation can support the creation of human-centered LLM applications: }designers effectively made use of the two-way translation between design requirements and their UX designs by using LLMs adapted with their requirements to improve interface affordances, while also using their designs to define additional requirements for model behavior. 
These approaches' promises were amplified once designerly adaptation was embraced as a collaborative practice. Our work illuminates paths for \replaced{designerly adaptation and its associated tools}{new processes and tools} to foreground designers' user-centered expertise for more responsible and thoughtful deployment of LLM-powered technologies.

\begin{acks}
We thank all our participants for their time and expertise, and reviewers for constructive feedback. We'd also like to thank members and interns of the FATE group at Microsoft Research for helpful comments and discussions, as well as Daniela Rosner for pointers to discourse on design codes.
\end{acks}

\bibliographystyle{ACM-Reference-Format}
\bibliography{refs}

\newpage
\appendix

\section{Formative Study Participants}
See Table \ref{t:participants-1}.
\label{a:formative}
\begin{table*}[h]
\centering
 \begin{tabular}{p{0.5cm} p{3cm} p{0.8cm} p{3.5cm} p{3.5cm} p{1.2cm}}
 \hline
 \bfseries P\# & \bfseries Job Title & \bfseries YoE & \bfseries LLM Application Area & \bfseries Education Background & \bfseries Region\\ 
 \hline
 P1 & UX Designer & 6--10 & Domain-specific QA & Visual/Industrial Design; Computing & Denmark\\ 
 P2 & UX Designer & 6--10 & Conversational Search & Visual/Industrial Design & Canada\\
 P3 & Principal UX Designer & 6--10 & Domain-specific QA & Visual/Industrial Design & U.S.\\
 P4 & Content Designer & 6--10 & Conversational Search; Domain-specific QA & Humanities & U.S.\\
 P5 & Principal Content Designer & 6--10 & Conversational Search & Humanities & U.S.\\
 P6 & Content Designer & 1--2 & Recommendation & Humanities & U.S.\\
 P7 & UX Researcher & 3--5 & Domain-specific QA & Visual/Industrial Design; Social \& Behavioral Sciences & Ireland\\
 P8 & UX Designer & 3--5 & Creativity Support Tools & Visual/Industrial Design & U.S.\\
 P9 & Senior UX Designer & 11+ & Conversational Search; Creativity Support Tools & Visual/Industrial Design & U.S.\\
 P10 & UX Researcher & 11+ & Domain-specific QA & Social \& Behavioral Sciences & U.S.\\
 P11 & UX Designer & 3--5 & Text Editing \& Generation; Domain-specific QA & Social \& Behavioral Sciences; Computing & U.S.\\
 P12 & Principal Content Designer & 11+ & Text Editing \& Generation; Domain-specific QA & Humanities & U.S.\\
 \hline
\end{tabular}

\caption{Details of our participants (job title, years of experience in design, LLM application area of their product/feature, and education background before starting their current role) in our formative study. All participants used Figma in their day-to-day work, and have experience designing user experiences for LLM-powered products.} 

\Description{A table summarizing the background information of 12 participants (P1 to P12) involved in our formative study, categorized by job title, years of experience (YoE), LLM application area, educational background, and region. Column 1: Participant Number (P#): Lists participant identifiers from P1 to P12. Column 2: Job Title: Includes titles such as UX Designer, Content Designer, Principal UX Designer, Principal Content Designer, and UX Researcher. Column 3: Years of Experience (YoE): Ranges from 1–2 years to 11+ years. Column 4: LLM Application Area: Lists areas such as Domain-specific QA, Conversational Search, Text Editing & Generation, Recommendation, Creativity Support Tools, and Domain-specific QA. Column 5: Educational Background: Includes fields like Visual/Industrial Design, Humanities, Computing, Social & Behavioral Sciences. Column 6: Region: Specifies the region of the participants, including U.S., Canada, Denmark, and Ireland.}
\label{t:participants-1}
\end{table*}

\section{\canvil{} System Prompt Template}
\label{a:prompt-template}
\lstset{basicstyle=\ttfamily,breaklines=true,breakindent=0pt,breakatwhitespace=true,frame=single}

\begin{lstlisting}
<|im\_start|>system

# Context
{Model profile field}

# Users
{Audience setting field}

# Core Instructions
{Core instructions field}

# Guardrails and Limitations
{Guardrails field}

# Example User Inputs and Responses
{Example inputs/outputs pairs}

# Final Instructions
You are the model described above. You will follow all instructions given to the model closely.

<|im\_end|>
\end{lstlisting}

\section{Design Study Recruitment Details}
\label{a:recruitment}
We recruited our participants from two channels. First, we distributed study invites to designers via email, professional interest groups, and word of mouth at a technology company. Some of these invitees took part in our formative study. We also snowball sampled by asking our invitees to refer us to other designers they work with who may also be interested in participating. From this channel, we only selected designers with prior experience working on LLM-powered products and features. Second, we sent study invites to a Slack workspace for HCI and design maintained by a large, public university in the United States. The population in the Slack consists primarily of design students and early-career designers. From this channel, we only selected those without any experience working on LLM-powered products and features. Our goal of recruiting from these two channels was to capture potential disparities that may arise in our findings hinging on prior experience working with LLMs. In the end, however, we did not notice meaningful qualitative differences in results between the two groups, except that those with LLM design experience drew more connections between adaptation and their past workflows for tinkering with LLMs or AI in general. 

We organized all candidates from the technology company into groups based on member preferences and availabilities, and did the same for candidates from the academic institution. In the end, we had 8 participants from the technology company and 9 from the academic institution.

\section{Design Study Participants}
\label{a:design}
See Table \ref{t:participants-2}.
\begin{table*}[h]
\centering
 \begin{tabular}{p{0.5cm} p{0.5cm} p{2.5cm} p{3cm} p{1cm} p{4cm}}
 \hline
 \bfseries G\# & \bfseries P\#& \bfseries Prior LLM Design Experience? & \bfseries Job Title & \bfseries YoE & \bfseries Education Background\\ 
 \hline
 \multirow{3}{=}{G1} & P1 & No & Design Student & 3--5 & Visual/Industrial Design\\ 
 & P2 & No & Design Student & 1--2 & Visual/Industrial Design\\
 & P3 & No & UX Designer & 3--5 & Visual/Industrial Design; Computing; Information\\
 \hline
 \multirow{3}{=}{G2} & P4 & No & UX Designer & 1--2 & Visual/Industrial Design; Computing\\ 
 & P5 & No & UX Designer & 3--5 & Visual/Industrial Design\\
 & P6 & No & UX Designer & 1--2 & Visual/Industrial Design\\
 \hline 
 \multirow{3}{=}{G3} & P7 & No & UX Designer & 1--2 & Visual/Industrial Design\\ 
 & P8 & No & UX Designer & 1--2 & Visual/Industrial Design; Management\\
 & P9 & No & Design Student & 1--2 & Visual/Industrial Design; Computing; Information\\
 \hline 
 \multirow{2}{=}{G4} & P10 & Yes & Principal Content Designer & 11+ & Humanities\\ 
 & P11 & Yes & Principal Content Designer & 6--10 & Humanities, Visual/Industrial Design; Social \& Behavioral Sciences\\
 \hline 
 \multirow{3}{=}{G5} & P12 & Yes & Senior UX Designer & 3--5 & Visual/Industrial Design\\ 
 & P13 & Yes & UX Designer & 3--5 & Visual/Industrial Design\\
 & P14 & Yes & UX Designer & 3--5 & Social \& Behavioral Sciences; Computing\\
 \hline 
 \multirow{3}{=}{G6} & P15 & Yes & Content Designer & 1--2 & Humanities\\ 
 & P16 & Yes & UX Researcher & 11+ & Social \& Behavioral Sciences\\
 & P17 & Yes & UX Designer & 1--2 & Information\\
 \hline
\end{tabular}

\caption{Details of our participants (job title, years of experience in design, LLM application area of their product/feature, and educational background before starting their current role) in our design study. All participants were based in the U.S., and used Figma in their day-to-day work.}
\Description{A table summarizing the background information of participants (P1 to P17) grouped into six groups (G1 to G6) in our design study, based on their prior LLM design experience, job title, years of experience (YoE), and educational background. Column 1: Group Number (G#): Participants are grouped from G1 to G6. Column 2: Participant Number (P#): Lists participant identifiers from P1 to P17. Column 3: Prior LLM Design Experience?: Indicates whether participants had prior experience designing LLMs ("Yes" or "No"). Column 4: Job Title: Includes titles like Design Student, UX Designer, Senior UX Designer, Content Designer, and Principal Content Designer. Column 5: Years of Experience (YoE): Ranges from 1–2 years to 11+ years. Column 6: Educational Background: Lists fields such as Visual/Industrial Design, Computing, Information, Social & Behavioral Sciences, Humanities, and Management.}
\label{t:participants-2}
\end{table*}

\section{Feasto User Groups}
\label{a:user-groups}

\subsection{Dimensions of Variance}
\begin{itemize}
    \item[\textbf{D1}] \textbf{Dietary restrictions}: users in different locations may have varying dietary restrictions due to religious and cultural customs (e.g., Halal in Turkey, vegetarianism in India).
    \item[\textbf{D2}] \textbf{Access to ingredients}: based on their region, users may have easier access to some ingredients than others.
    \item[\textbf{D3}] \textbf{Menu style preferences}: some cultures may prefer precise and detailed menus and recipes, while others prefer looser guidelines or drawing from traditional cooking techniques. 
\end{itemize}

\subsection{User Group Descriptions}
\subsubsection{North America}
A prominent percentage of users in North America live in the state of California. They enjoy a wide range of cuisines but many also consider themselves as health- and environment-conscious, adopting pescatarian diets to avoid heavy consumption of meat. Some have also reported nut allergies. 

These users like to take advantage of the fresh fruits and vegetables that grow in the California sun, such as spinach, kale, citrus fruits, avocados, peaches, and berries. Due to their proximity to the ocean, many also enjoy diverse varieties of seafood. 

When preparing food, these users tend to follow precise recipes.

\subsubsection{Middle East}
Turkey is home to many users from the Middle East. Because of the country’s predominantly Muslim population, many users follow the Islamic dietary laws (halal) and enjoy traditional Mediterranean cuisine. They avoid pork and alcohol and also fast during the month of Ramadan, during which they eat only during specific hours. 

These users enjoy access to a variety of Mediterranean ingredients like lamb, poultry, pita bread, olives, dates, and a range of spices like paprika, coriander, and cinnamon. 

When preparing food, these users tend to treat recipes as loose guidelines rather than precise instructions.

\subsubsection{Asia}
A significant portion of users in Asia hail from India. Many of these users adhere to a vegetarian diet due to their Hindu beliefs, while others may consume white meat such as poultry or fish but rarely beef. Many also avoid onion and garlic during certain religious festivals. 

These users enjoy access to a wide range of South Asian ingredients such as lentils, chickpeas, paneer (cheese); various earthy spices such as turmeric, cumin, and cardamom; and tropical fruits including papayas, guava, and mangoes.

These users rarely use written recipes, but instead rely on recipes and techniques passed on through word of mouth.

\end{document}